\documentstyle[12pt]{article}
\input epsf.tex

\newcommand{\be}{\begin{equation}}
\newcommand{\ee}{\end{equation}}
\newcommand{\bea}{\begin{eqnarray}}
\newcommand{\eea}{\end{eqnarray}}
\newcommand{\bd}{\begin{displaymath}}
\newcommand{\ed}{\end{displaymath}}

\def\ie{{\it i.e.}}
\def\diag{\mbox{diag}}
\def\dim{\mbox{dim}}
\def\End{\mbox{End}}
\def\Endskew{\mbox{Endskew}}
\def\Hom{\mbox{Hom}}
\def\tr{\mbox{tr}}
\def\Ker{\mbox{Ker}}
\def\Tr{\mbox{Tr}}
\def\cd{{\cal D}}
\def\ct{{\cal T}}

\def\o{{\scriptscriptstyle 0}}
\def\G{\Gamma}
\def\W{\Omega}
\def\a{\alpha}
\def\b{\beta}
\def\g{\gamma}
\def\d{\delta}
\def\eps{\epsilon}

\def\l{\lambda}
\def\m{\mu}
\def\n{\nu}
\def\r{\rho}
\def\s{\sigma}

\def\u{\vartheta}
\def\vv{\nu}
\def\z{\zeta}
\def\w{\psi}
\def\mn{\mu\nu}

\def\xbar{\bar x}

\def\wt{\widetilde}
\def\de{\partial}

\def\mezzo{{1 \over 2}}
\def\unoar{1-({a \over r})^4}

\def\ALE{Asymptotically~Locally~Euclidean~}

\def\YM{Yang-Mills~}

\def\KN{Kronheimer-Nakajima~}
\def\Kro{Kronheimer~}
\def\Nak{Nakajima~}
\def\GH{Gibbons-Hawking~}

\def\tb{tautological~bundle~}

\def\zm{zero-mode~}
\def\zms{zero-modes~}

\def\hkq{hyper-K\"ahler~quotient~}
\def\hk{hyper-K\"ahler~}

\def\sd{self-dual~}
\def\asd{anti-self-dual~}

\def\stand{standard~embedding~}
\def\min{minimal~instanton~}
\def\mc{multicenter~}
\def\mom{moment~map~}
\def\unita{{1 \kern-.30em 1}}
\def\fey{{\big / \kern-.80em D}}
\font\mybb=msbm10 at 12pt
\font\mybbb=msbm10 at 8pt
\def\bb#1{\hbox{\mybb#1}}
\def\bbb#1{\hbox{\mybbb#1}}

\def\complex{{\bb{C}}}
\def\ccomplex{{\bbb{C}}}
\def\zet{{\bb{Z}}}
\def\zzet{{\bbb{Z}}}
\def\real{{\bb{R}}}
\def\rreal{{\bbb{R}}}

\def\R4{\real^4}
\begin{document}

\thispagestyle{empty}
\begin{flushright}
ROM2F-95-30
\end{flushright}
\centerline{\large \bf Explicit Construction of} 
\centerline{\large \bf Yang-Mills Instantons on ALE Spaces}
\vspace{0.2cm}
\centerline{\bf Massimo Bianchi, Francesco Fucito, Giancarlo Rossi}
\vskip 0.1cm
\centerline{\sl Dipartimento di Fisica, 
Universit\`a di Roma II ``Tor Vergata"}  
\centerline{\sl I.N.F.N. \ - \  Sezione di Roma II, }
\centerline{\sl Via Della Ricerca Scientifica \ \ 00133 \ Roma \ \ ITALY}
\vskip .2cm
\centerline{{\bf Maurizio Martellini} 
\footnote{On leave of absence from 
Dipartimento di Fisica, Universit\`a di Milano, 20133 Milano, Italy}}
\vskip 0.1cm
\centerline{\sl The Niels Bohr Institute, University of Copenhagen}
\centerline{\sl DK-2100 Copenhagen $\phi$, Denmark}
\centerline{\sl I.N.F.N. - Sezione di Milano} 
\centerline{\sl Landau Network at ``Centro Volta", Como ITALY}
\vskip .3cm
\centerline{\large \bf ABSTRACT}
{We describe the explicit construction of \YM instantons on
\ALE (ALE) spaces, following the work of Kronheimer and Nakajima. 
For multicenter ALE metrics, we determine the abelian
instanton connections which are needed for the
construction in the non-abelian case. We compute the partition
function of Maxwell theories on ALE manifolds and comment on
the issue of electromagnetic duality. We discuss the topological
characterization of the instanton bundles as well as the identification
of their moduli spaces. We generalize the 't Hooft ansatz
to $SU(2)$ instantons on ALE spaces and on other \hk manifolds. 
Specializing to the Eguchi-Hanson gravitational background, we 
explicitly solve the ADHM equations for $SU(2)$ 
gauge bundles with second Chern class 1/2, 1 and 3/2.}
\newpage

\setcounter{section}{0}
\section{Introduction}
\setcounter{equation}{0}
\ALE (ALE) gravitational instantons have played a fundamental role in 
Euclidean quantum
supergravity (for a review see \cite{egh}). They induce non-perturbative 
effects which turn out to be very relevant for understanding the vacuum
structure of the theory and are possibly responsible for the emergence of
dynamical breaking mechanisms for supersymmetry (SUSY). 
From a mathematical point of
view they are interesting non-compact \hk manifolds which admit a 
complete classification, induced by the A-D-E classification of
simply-connected Lie algebras.

In a recent work \cite{bfmr} we have shown that ALE metrics can be 
incorporated
in a fully supersymmetric solution of the four-dimensional heterotic 
string equations
of  motion
with constant dilaton and zero torsion (for a review of SUSY solutions of low 
energy supergravities see \cite{dkl}). Thanks to the so-called
\stand condition, this solution can be shown to be an exact one
(corresponding to an exact $N=(4,4)$ superconformal field theory on the 
world-sheet, see also 
\cite{abfgz, kkl}), to all orders in (the $\sigma$-model 
coupling) $\a^{\prime}$.

To study whether these backgrounds might eventually lead to the 
the formation 
of chiral condensates, instanton dominated correlation functions
must be computed. For globally supersymmetric \YM (YM) theories with 
$N=1, 2, 4$ we have performed this computation 
around the minimal instanton on the Eguchi-Hanson (EH) manifold 
\cite{bbfmr}, 
the simplest of all possible ALE spaces.  The denomination minimal 
instanton comes
from the value of its second Chern class, $\kappa={1/2}$, which 
is the minimal
one allowed for $SU(2)$ bundles on the EH manifold. 
The minimal instanton plays on the EH background 
the same role as the t'Hooft-Polyakov instanton on flat 
space. It was also found, by quite different 
methods, in \cite{bcc}. 
The results of the instantonic computations presented in \cite{bbfmr} are 
analogous to those obtained in flat spacetime \cite{akmrv}: 
the chiral condensates are constant and turn out to be
proportional to the appropriate power of the renormalization group 
invariant scale of the theory, as dictated by naive dimensional counting.

The extension of instanton computations to fullfledged solutions  of 
the heterotic string equations of motion \cite{bfmr}, 
is not straightforward. 
It requires, as we said, the embedding of the spin connection into the 
gauge group, and it is based on the fact that the self-dual spin connection 
of an ALE gravitational background lies in the
$SU(2)_L$ factor of the $SO(4)$ Lorentz group. The relevant 
$SU(2)$ gauge bundle is then the holomorphic tangent bundle to the ALE
manifold. In the case of the EH background the dimension of the moduli 
space of
the gauge connections on the tangent bundle was computed, via index 
formulae, to
be $12$ and its second Chern class turned out to be 
$\kappa={3/2}$ \cite{bfmr}.
Computation of instanton  effects requires the knowledge of the form of
the gauge connection. This is the task  we tackle in the present paper 
where we 
build all $SU(2)$ gauge bundles up to $\kappa={3/2}$ and 
clarify the quite intricate mathematics needed to do it.
To carry out this program we will use recent results of Kronheimer 
and Nakajima who, in a series of beautiful 
papers \cite{kn, nak, gn}, completely solved the problem of 
constructing self-dual YM connections on  ALE manifolds 
along the lines of the analogous 
Atiyah-Drinfeld-Hitchin-Manin (ADHM) construction \cite{adhm, aty}
of gauge instantons on flat space.

The plan of the paper is as follows. In section 2 we review 
the construction of ALE gravitational instantons. 
Using twistor techniques, Hitchin \cite{hitb} has shown that ALE 
manifolds are 
smooth resolutions of algebraic varieties in $\complex^3$. 
Simple singularities
admit an A-D-E classification according to which, the class of \mc 
metrics of \GH (GH) \cite{gh}
(to which the EH instanton \cite{eh} belongs) may be 
identified with the resolution of singularities of A-type.
A general construction of all ALE manifolds was then worked out by
Kronheimer \cite{kro}. These manifolds emerge
as minimal resolutions of $\complex^2/\Gamma$, where the discrete 
subgroups $\Gamma$ of $SU(2)$ are 
in one-to-one correspondence 
with the extended Dynkin diagrams of simply-laced (\ie~A-D-E) simple Lie 
algebras. ALE spaces are explicitly
obtained as \hk quotients of flat Euclidean spaces. This
allows, in principle, the determination of the \hk metric on them.
In practice only for ALE spaces of A-type, corresponding to $\G=\zet_N$,
the metric is known and it turns out to be diffeomorphic to the GH 
\mc metric. 

The \hk quotient construction allows to identify a principal bundle 
over
the ALE space whose natural connection has \asd curvature. 
To this principal bundle, a tautological vector bundle, $\ct$, can 
be associated  
by a change of fiber. This bundle admits a decomposition under
the action of $\G$ into $r$ (=rank($\G$)) elementary 
bundles, $\ct_i$, whose 
first Chern classes form a basis for the second cohomology group of 
the ALE space. For \mc ALE metrics the bundles $\ct_i$ admit
abelian connections that we explicitly construct in section 3. 
They will be needed in the following to compute the 
gauge connections in non-abelian cases. 
As a byproduct we obtain a formula for the partition function
of abelian gauge theories on ALE spaces, in terms of the level one 
characters
of the affine Lie algebra of the type A-D-E, associated to the ALE
space under consideration, and we show that electromagnetic duality, 
at least in
its simplest form, does not hold on ALE spaces.

In Section 4 we recall the essential steps of the ADHM construction of
gauge instantons on $\real^4$ and argue that this construction can 
be viewed
as an \hk quotient. This observation establishes a bridge with 
the \KN (KN) construction of YM instantons
on ALE spaces which is discussed immediately afterwards.
We then elaborate on some known solutions, 
generalizing the ansatze of 't Hooft \cite{thooft} and Jackiw, Nohl 
and Rebbi 
\cite{jnr} to four-dimensional \hk manifolds (including ALE
gravitational instantons) and discuss some $\G$-invariant 
instantons on $\R4$.

In section 5 by studying the topological properties of the gauge 
instanton
bundles, we are able to identify some already known solutions.
We emphasize the role of the discrete group $\G$ and
compute the topological invariants of these bundles, 
the dimensions of their moduli spaces and, in the $SU(2)$ case,
the indices of the relevant Dirac operators.

In section 6 we explicitly solve the ADHM equations for $SU(2)$
gauge bundles with
second Chern class up to $\kappa={3/2}$ on the EH background.
Following the KN-ADHM construction the problem is reduced to 
a bunch of conceptually simple but somewhat tedious algebraic manipulations.

In section 7 we discuss some properties of the moduli spaces
of YM instanton connections on ALE spaces and exhibit the \hk metric for 
the minimal instanton on the EH background and in section 8 we draw our
conclusions.

In the appendix we show how to recover the explicit form of 
the EH metric, following the \hk quotient construction.

\section{\bf Kronheimer Construction of ALE Instantons}
\setcounter{equation}{0}
\subsection{ Hyper-K\"ahler Quotients}

A Riemannian manifold $X$ is said to be \hk if it is equipped
with three covariantly constant complex structures $I, J, K$, 
\ie~automorphisms of the tangent bundle satisfying the 
quaternionic algebra $I^2=J^2=K^2=-\unita$ and $IJ=-K,JK=-I,KI=-J$.
In this circumstance the metric $g$ on $X$ is said to be of K\"ahler type
(or simply, K\"ahler) with respect to $I,J,K$ and one can define three closed
K\"ahler 2-forms 
\bea
\omega_I (V,W) &=& g(V,IW)\cr    
\omega_J (V,W) &=& g(V,JW)\cr   
\omega_K (V,W) &=& g(V,KW)
\label{dueuno}
\eea
with $V, W$ vector fields on $X$. 
We will often denote the \hk forms in (\ref{dueuno}) as $\omega_i$,
with $i=I,J,K$. In four dimensions a simply-connected Riemannian manifold
is \hk when its Riemann curvature tensor is either self-dual or 
anti-self-dual\footnote{In the mathematical literature, 
instantons on complex manifolds denote YM connections  
whose field-strength is anti-self-dual in the natural orientation of the 
manifold, inherited from its
complex structure. In these conventions,
(ALE) gravitational instantons, which are complex \hk manifolds, 
turn out to have anti-self-dual curvature.
We will follow, instead, the physicists' conventions of inverting the
natural orientation and call instantons, either YM or gravitational, 
connections with self-dual curvature.}.
A four-dimensional \hk manifold is thus a gravitational instanton. 

In this section we want to briefly recall the \Kro \cite{kro} 
construction of a
particular family  of \hk manifolds, the so called ALE 
gravitational instantons.

We begin by discussing a method to produce a \hk manifold $X$
of real dimensions $d=4(n-k)$ 
starting from a $4n$-dimensional \hk manifold $\Xi$ \cite{hklr}.
The manifold $X$ is obtained as the quotient of a real
subspace of $\Xi$ by some subgroup of the isometry group of $\Xi$.
In this procedure a central role is played by the moment map. 
By assumption $\Xi$ admits vector fields $V$, known as Killing vectors, 
which generate isometries, \ie~for which ${\cal L}_V g = 0$,
with ${\cal L}_V$ the Lie derivative along $V$.
Killing vector fields are said to be triholomorphic if 
furthermore ${\cal L}_V \omega_i = i_V d\omega_i + d(i_V \omega_i) = 0$,
where $i_V$ denotes contraction with $V$.
To each vector of this kind there correspond three Killing potentials,
$\m_i^V$, which can be obtained by integrating the equations
\be 
d\m_i^V = i_V \omega_i 
\label{duedue}
\ee
following from $d\omega_i=0$ and ${\cal L}_V \omega_i = 0$.
In the absence of abelian factors, the integration constants may 
be fixed requiring $W\m_i^V = \m_i^{[V,W]}$. The 
Killing potentials $\{\m_i^V\} =\m_i $ define a \hk \mom  
\be
\mu_i\,\colon\quad \xi\in \Xi \mapsto \m^a_i(\xi)\in\real^3 
\times {\cal G}^* 
\quad i=1,2,3\quad a=1,\ldots,\dim(G)
\label{mommap}
\ee 
where 
${\cal G}^*$ is the dual to the Lie algebra, ${\cal G}$, of the 
isometry group 
$G$ generated 
by the triholomorphic vectors, $V$. The moment maps in (\ref{mommap}) can be 
suggestively reorganized in the combinations
$\m_\rreal=\m_3$ and $\m_\ccomplex=\m_1 +i \m_2$.

Let us make an example to clarify the 
meaning of these concepts. In Hamiltonian mechanics, there is a 
natural two form:
\be
\omega=dp^i\wedge dq^i
\label{duedueuno}
\ee
where the $q^i$'s and the $p^i$'s are the generalized coordinates and 
momenta. If we take the isometries to be translations, 
$V=a^i{\de\over\de q^i}$, then
$i_V\omega=a^idp^i=d\mu$ and $\mu$ is the usual linear momentum, 
whence the name. If the $V$'s were the generators 
of the rotation group, $SO(3)$, then the moment map 
would be given by $\vec\mu=\vec q\wedge\vec p$, \ie~by the angular
momentum.  It should be clear now that the \mom construction is a way of 
generalizing the procedure of identifying conserved quantities 
in classical mechanics.

Any \hk manifold $\Xi$ admitting a compact group $G$ of $k$ freely 
acting
triholomorphic isometries, contains a \hk submanifold $X$ of real dimension
\be
\dim(X) = \dim(\Xi) - 4 \dim(G) = 4n - 4k
\label{duetre}
\ee
which is the \hk quotient of $\Xi$ with respect to $G$.
The construction of $X$ proceeds in two steps.
First a submanifold $P_{\z}$ of dimension $\dim(P_{\z}) = \dim(\Xi) - 3\dim(G) 
= 4n - 3k$, is identified as the level set of $3\times k$ \hk 
moment maps, \ie
\be
P_{\z} = \{\xi\in \Xi \, \vert \; \m_i^a(\xi) = \z_i^a, \quad i=1,2,3 \quad 
a=1,\dots, k\}
\label{duequattro}
\ee
When the $\z$'s belong to $\real^3 \times {\cal Z}^*$, with ${\cal Z}^*$
the dual to the center of ${\cal G}$, the 
hypersurface $P_{\z}$ is left invariant by the action of $G$. 
In fact $P_{\z}$ turns out to be a $G$-principal bundle.
If one divides out the group $G$ from $P_{\z}$, one obtains a new \hk
manifold, $X_{\z}=P_{\z}/G$. Notice that the algebro-geometric quotient
$N_{\z}/G^\ccomplex$, with 
\be
N_{\z} = \{\xi\in \Xi  \, \vert \; \m_{\ccomplex}^a(\xi) = 
\z_{\ccomplex}^a, \quad 
a=1,\dots, k\}
\label{algeom}
\ee
and $G^\ccomplex$ the complexification of $G$, is diffeomorphic 
to $X_{\z}$.
As a by-product of this construction one finds that 
the curvature of the natural connection on the principal bundle
$P_\z$ over $X_\z$ is \sd \cite{hklr,gn}.  

\subsection{A-D-E Classification of ALE Gravitational Instantons}

\Kro constructed all ALE gravitational instantons as particular 
\hk quotients \cite{kro}. Given a discrete Kleinian subgroup of SU(2), 
\ie~$\G = \zet_N,D_N^*,O^*,T^*,I^*$, consider the flat
\hk space 
\be
\Xi=({Q\otimes {\End}(R)})_{\G}
\label{duecinque}
\ee
where $Q (\sim\complex^2)$ is the vector space of the fundamental 
two-dimensional
representation, $\rho_{_Q}$, of $\G$ and ${\End}(R)$ is the set of
endomorphisms of the vector space, $R$, of the regular representation,
$\rho_{_R}$, of $\G$. $\Xi$ is the space of $\G$-invariant pairs of
endomorphisms of $R$. Since for the regular representation 
$\dim(R) = \dim(\G) =
|\G|$, the elements of $\Xi$ can be represented as doublets of the form 
\be 
\xi=\pmatrix{\a\cr \b\cr} 
\label{punti}
\ee
where $\a,\b$ are $|\G|\times|\G|$ complex matrices satisfying the 
$\G$-invariance property\footnote{With an abuse of notations we 
will sometimes 
call the vector space, $R$, that carries the representation 
$\rho_{_R}$, the
``representation $R$" and use the symbol $R$ also for the 
representation
matrix, $\rho_{_R}$.} 
\be 
\pmatrix{\rho_{_R}(\g) \a \rho_{_R}(\g^{-1}) \cr
\rho_{_R}(\g) \b \rho_{_R}(\g^{-1})\cr} = \rho_{_Q}(\g) 
\pmatrix{\a\cr \b\cr}
\quad \gamma\in\G
\label{gamminva}
\ee
The points $\xi$ of the manifold $\Xi$ can also be represented by  
quaternions of matrices  
\be
\xi= \pmatrix{\a&-\b^\dagger\cr \b &\a^\dagger\cr}
\label{duedieci}
\ee  
so that $\Xi$ can also be viewed as the space $\Xi =(T^*\otimes_\rreal 
{\Endskew}(R))_\G$, with $T$ the quaternionic space and $T^*$ its dual.

The role of the discrete subgroup $\G\subset SU(2)$ in (\ref{gamminva}),
is clarified by the Mac Kay correspondence \cite{mkay} between Kleinian 
subgroups
$\G\subset SU(2)$ and A-D-E extended Dynkin diagrams, $\wt{\Delta}_\G$. 

We recall that, given a representation $\rho_{_W}$ of $\G$, 
in order to find 
its decomposition, $\rho_{_W}=\oplus_{i=0}^{r-1} w_i \rho_i$, 
in irreducible representations of $\G$ ($\rho_{_R{_i}}\equiv \rho_i$, 
where $\rho_o$ is the trivial representation), we must look at the
extended Dynkin diagram, $\wt{\Delta}_{\G}$, of the A-D-E Lie algebra
associated to $\G$. In the above decomposition $r$ is the rank of 
$\G$, \ie~the
number of conjugacy classes of $\G$. The extended Dynkin diagram is 
constructed, starting from the standard
Dynkin diagram, $\Delta_{\G}$, according to the following steps 
\begin{itemize}
\item [{i)}] an extra dot,
corresponding to minus the highest root of the algebra 
(called the extended root, $\a_{\o}$),
is added to $\Delta_{\G}$; 
\item [{ii)}] a number, $n_i$, equal to half the sum of the
numbers attributed to the neighbouring roots is associated to each dot of
$\wt{\Delta}_{\G}$, starting from the assignment $n_{\o}=1$ made to the
extended root. 
\end{itemize}
In this way one finds
$\a_{\o}=-\sum_{i=1}^{r-1}\,n_i\a_i$ and the numbers $n_i$ turn out 
to coincide
with the dimensions of the irreducible representations $\rho_i$ of 
$\G$. In the
case $\G=\zet_N$, the extended Dynkin diagram corresponds to that 
of the Lie
algebra of $SU(N)$ (the algebra $A_{N-1}$ in the A-D-E classification), 
with all $n_i=1$, as shown in Fig.1. As a consequence,
$\rho_{_W}$ decomposes in one-dimensional representations. 

\vskip .6cm
\centerline{\vbox{\epsfysize=40mm \epsfbox{dynkin.eps}}}
\smallskip
\vskip .4cm
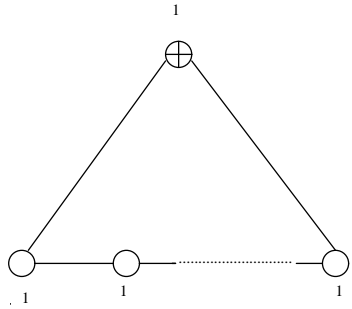
\centerline{\bf Figure 1}
\vskip .6cm

Denoting by $R_i$ the space of the irreducible $n_i$-dimensional 
representations
of $\G$, the decomposition 
\be
Q\otimes R_i = \oplus_j A_{ij} R_j
\label{duesei}
\ee
makes the correspondence between Kleinian subgroups
$\G\subset SU(2)$ and A-D-E extended Dynkin diagrams quite 
explicit, as it
turns out that $A_{ij} = 2 \d_{ij} - \tilde C_{ij}$ with 
$\tilde C_{ij}$ the 
extended Cartan matrix of $\wt{\Delta}_\G$ \cite{slan}. 
For later use we remark here
that the vector $n=(n_{\o}, n_1,\dots, n_{r-1})$, whose components 
are the
dimensions of the irreducible representations of $\G$, belongs to 
the null space
of the extended Cartan matrix, \ie~$\sum_{j=0}^{r-1}\tilde
C_{ij}n_j=0$ {\footnote{From now on, unless differently stated, 
all the sums and
products over indices labelling the irreducible representations 
of $\G$ are
understood to be extended from 0 to $r-1$.}.

The vector space carrying the regular representation admits
the decomposition
\be 
R=\oplus_i R_i \otimes \bar R_{i}
\label{duesette}
\ee 
In the following we will be interested in the action of $\G$ by left 
multiplication only. Since left multiplication leaves $\bar R_i$ 
untouched, one
can simply write $R=\oplus_i R_i \otimes \complex^{n_i}$.
In this respect $\G$ can be taken to be a subgroup of the same 
$SU(2)_L$ factor of the Lorentz group
in which the self-dual spin connection of the resulting ALE 
instanton will lie. 

Imposing $\G$-invariance in the form of (\ref{gamminva}), 
one finds the decomposition
\be
\Xi=\oplus_{ij} A_{ij} \Hom(\complex^{n_i}, \complex^{n_j})
\label{dueotto}
\ee
which shows that the (real) dimension of $\Xi$ is 
\be
\dim(\Xi) = 2 \sum_{i,j=0}^{r-1} A_{ij} n_i n_j = 4 
\sum_{i=0}^{r-1} (n_i)^2
= 4 | \Gamma |
\label{duenove}
\ee
To prove (\ref{dueotto}) we write successively
\bea
\Xi &=&({Q\otimes {\End}(R)})_{\G}=
(Q\otimes\Hom(\oplus_i R_i\otimes\complex^{n_i}, 
\oplus_j R_j\otimes\complex^{n_j}))_{\G}\cr
&=&(Q\otimes(\oplus_{ij} \Hom(R_i,R_j)))_{\G}\otimes\,
\Hom(\complex^{n_i},\complex^{n_j})\cr
&=&(\oplus_{ijk} A_{ik} \Hom(R_k,R_j))_{\G}\otimes\,
\Hom(\complex^{n_i},
\complex^{n_j})
\label{dueottobis}
\eea
having used the decomposition (\ref{duesei}). 
Since $\Hom(R_i,R_j)_{\G}=\delta_{ij}$ thanks to Schur's lemma, 
equation (\ref{dueotto}) immediately follows.

Using the representation (\ref{duedieci}), 
the three \hk forms on $\Xi$ can be written
\be
\omega_i = \Tr (d\xi^\dagger\wedge d\xi \s_i^P) \quad i=1,2,3
\label{dueundici}
\ee
where the $\s_i^P$'s are the standard Pauli matrices. 
They can be combined into a
real $(1,1)$-form and a complex $(2,0)$-form as follows
\bea
\omega_\rreal &=&\Tr (d\a\wedge d\a^\dagger)+\Tr (d\b\wedge d\b^\dagger)\cr
\omega_\ccomplex &=&\Tr (d\a\wedge d\b)
\label{duedodici}
\eea
The action of a generic elements $g\in U(|\G|)$ on $\xi\in \Xi$
is induced by the transformations
\be
\a \rightarrow g \a g^\dagger,  \quad 
\b \rightarrow g \b g^\dagger  
\label{duedodicia}
\ee
and leaves invariant the \hk forms (\ref{duedodici}) and the flat 
metric on $\Xi$.
The representation  space $R$ is naturally acted 
upon by $U(|\G|)$. Requiring $\G$-invariance on $R$ 
reduces $U(|\G|)$ to $G'=\otimes_{i=0}^{r-1} U(n_i)$.
The group  of freely acting triholomorphic 
isometries of $\Xi$ that will be used to perform the \hkq is 
$G=\otimes_{i=1}^{r-1} U(n_i)$, where the factor $U(n_0)=U(1)$, 
acting trivially on $\Xi$, has been eliminated.
Using the invariance of (\ref{duedodici}) under the transformation 
(\ref{duedodicia}), one can construct the \hk moment maps 
which turn out to be \cite{kro}
\bea
\mu_\ccomplex &=&[\a,\b]\cr
\mu_\rreal &=&[\a,\a^\dagger]+[\b,\b^\dagger]
\label{duetredici}
\eea
The level sets, $P_{\z}$ and $N_{\z}$, defined 
in (\ref{duequattro}) and (\ref{algeom}), can be explicitly identified with
\bea
[\a,\b] &=& \z_\ccomplex \cr
[\a,\a^\dagger]+[\b,\b^\dagger]&=&\z_\rreal
\label{duetredicibb}
\eea
where $\z\in \real^3\otimes{\cal Z}^*$, with 
${\cal Z}^*$ the dual to the center of the Lie algebra of 
$G$, \ie~$\z = \oplus_{i=0}^{r-1} \z_i \unita_{n_i}$ with 
$\sum_{i=0}^{r-1} \z_i=0$.
The last step of the \hk quotient can be 
eventually performed and the resulting space $X_{\z}=P_\z/G$ turns 
out to have
dimension 
$\dim(X_{\z})=\dim(\Xi)-4\dim(G)=4|\G|-4(|\G|-1)= 4$. For generic values of 
the deformation parameters, $\z$, $X_\z$ is a smooth \hk manifold. 
In the final form of the metric the parameters $\z_\ccomplex$, 
which describe the deformations of the complex structure, can be 
reabsorbed through a non-analytic change of coordinates on $X$, 
and the choice $\z_\ccomplex=0$ 
can be made without loosing generality, much as in the case of gauge 
instanton connections where the parameters
associated to global gauge transformations are not explicitly displayed. 

On the contrary, setting the deformation parameters $\z_\rreal=0$, we 
get the orbifold $X_0=\complex^2/\G$. 
\Kro has shown \cite{kro} that every ALE \hk four-manifold $X_\z$
is diffeomorphic to the minimal resolution of $X_0$, thus 
proving the completeness of the construction we have just described.

An explicit example of the \hkq construction is given in the appendix where 
the EH metric is derived.

\section{Tautological Bundle and Abelian Instantons}
\setcounter{equation}{0}
As already mentioned, the principal bundle $P_\z$ admits a natural 
connection with \sd curvature. By a change of fiber, a 
tautological vector bundle $\ct$ with
typical fiber $R$ can be associated to $P_\z$. This vector bundle is 
tautological in the sense that (see (\ref{duecinque})) 
the points of
the base manifold, $X$, are endomorphisms of $R$ itself.
Under the action of $\G$, the tautological bundle $\ct$ admits the 
decomposition: $\ct =\oplus_i \ct_i\otimes \bar R_i$.
Apart from the trivial bundle $\ct_{_0}$ associated to the trivial 
(one-dimensional) representation of $\G$, 
the elementary bundles $\ct_i$ admit \sd connections which
are asymptotic to flat connections with holonomy determined by the
representation $\rho_i$ \cite{kn}.
The first Chern classes, $c_1(\ct_i)$, $i\neq 0$ ($c_1(\ct_{_0})=0$), form a
basis of the cohomology group, $H^2(X,\real)$, and satisfy \cite{kn} 
\be
\int_X c_1(\ct_i) \wedge c_1(\ct_j) = (C^{-1})_{ij} \qquad i,j=1,\dots,r-1
\label{intertau}
\ee
where $C^{-1}$ is the inverse of the Cartan matrix of 
the unextended Dynkin diagram, $\Delta_{\G}$.
The dual homology basis of $H_2(X,\zet)$ consists of the two-spheres,
$\Sigma_i$, which arise from the resolution of the exceptional set in 
$\ccomplex^2/\G$ \cite{hitb, kro}.

ALE manifolds associated to the discrete groups, $\G=\zet_N$, \ie~to the 
Dynkin diagrams of Lie algebras of $A_{N-1}$ type,
admit the \mc (self-dual) metrics \cite{hitb, gh} 
\be
ds^2 = V^{-1}({\vec x}) (dt +
{\vec\omega}\cdot d {\vec x})^2 + V({\vec x})d{\vec x}\cdot d {\vec x}
\label{multicenter}
\ee 
with
\be
V({\vec x}) =\sum_{i=1}^{N}
{1\over\mid {\vec x}- {\vec x}_i\mid}
\label{potv}
\ee
The functions $V({\vec x})$ represent the localized solutions of the equation 
$\nabla^2 V = 0$, which follows from the self-duality condition on the spin
connection 
\be 
{\vec \nabla}V={\vec\nabla}\times{\vec\omega}
\label{nablav}
\ee
When, as in this case, the base manifold, $X$, is a \mc ALE gravitational 
instanton,
the explicit expression of the \sd curvatures of $\ct_i$ may be found
using the following argument. 
The $N-1$ minimal surfaces, $\Sigma_i$, homologically equivalent 
to the uncontractible two-spheres, mentioned above, can be taken to be 
\cite{hitb}
\be
\Sigma_i = \{(\vec x,t)\in X: 
\vec x = \vec x_i + \lambda(\vec x_{i+1}-\vec x_i); 
\lambda\in(0,1), t\in(0,4\pi) \}
\label{surface}
\ee 
The intersection form of these $N-1$ two-spheres is given by (minus)
the Cartan matrix $C$ of $A_{N-1}$.
By identifying the periods of the three K\"ahler forms along 
the $\Sigma_i$
with the moduli 
of the
multicenter metrics (\ref{multicenter}),  
the latter can be shown
to coincide with the positions of the centers, $\vec x_i$ \cite{hitb, kro}. 
An explicit basis for $H^2(X,\real)$, consisting of $N$ \sd
two-forms, may be found starting from the ansatz \cite{yui} 
\be
F = E_l({\vec x})(e^\o\wedge e^l + \mezzo \eps^l{}_{mn} e^m \wedge e^n) 
\quad l,m,n=1,2,3
\label{frametwo}
\ee
with $E_l({\vec x})$ functions of ${\vec x}$ to be determined 
by the closure condition $dF=0$. If we choose the tetrad
corresponding to (\ref{multicenter}) to be 
\be
e^\o = e^\o_\m dx^\m = V^{-\mezzo} (dt + 
{\vec\omega}\cdot d{\vec x}),\quad\quad
e^l = e^l_\m dx^\m = V^{\mezzo} dx^l
\label{frame}
\ee
the two-forms (\ref{frametwo}) become
\be
F = E_l({\vec x}) B_l  
\label{twoforms}
\ee
with
\be
B_l = (dt + {\vec\omega}\cdot d{\vec x})\wedge dx^l +
\mezzo \eps^l{}_{mn} V dx^m \wedge dx^n
\label{twoformsdef}
\ee
Imposing the closure condition, one finds the $N$ solutions
$E_l^{(i)} = \nabla_l(V^{(i)}/V)$, where $V^{(i)} = {1/|x-x_i|}$ and 
$\vec\nabla V^{(i)} = \vec\nabla \times \vec\omega^{(i)}$. 
From the definition of $V$ in (\ref{potv}),
it is clear that $\sum_i E_l^{(i)}=0$, so that only $N-1$ out of the $N$
two-forms, $F^{(i)} = E_l^{(i)}({\vec x}) B_l$ are linearly independent. 
A convenient choice of basis is
${\cal F}^i = {1\over 4\pi} (F^{(i)} - F^{(i+1)})$ for $i=1,\cdots,N-1$.
In order to relate ${\cal F}^i$ to $c_1(\ct_i)$,
which are dual to the chosen homology basis in (\ref{surface}), \ie~satisfy
$\int_{\Sigma_j} c_1(\ct_i) = \delta_i{}^j$, it is enough to note that
\be
\int_{\Sigma_j} {\cal F}^i = C^{ij} 
\quad \int_X {\cal F}^{i} \wedge {\cal F}^{j} = C^{ij}
\label{intersection}
\ee
Comparing (\ref{intertau}) with (\ref{intersection}), one 
concludes that the relation between the two bases $\{ {\cal F}^i\}$ and 
$\{ c_1(\ct_i)\}$ is 
\be
c_1(\ct_i) = \sum_j(C^{-1})_{ij} {\cal F}^j   
\label{baschange}
\ee
In the KN construction of YM instantons on $X$,
which will be described in the next section, it will prove convenient to
introduce the monopole (or better the abelian instanton) potentials 
\be
A^i = A^i_\m dx^\m = {1\over 4\pi} V^{-\mezzo} ((V^{(i)} - V^{(i+1)}) e^\o + 
(\vec\omega^{(i)} - \vec\omega^{(i+1)})\cdot\vec{e})
\label{monopot}
\ee
such that locally ${\cal F}^i = dA^i$.
Using (\ref{monopot}), it will be similarly possible to write
the \sd two forms, $c_1(\ct_i)$, in terms of the monopole potentials 
$A^{\ct_i} = (C^{-1})_{ij} A^j$.

We now wish to make a brief detour on abelian gauge theory 
and discuss some issues concerning electromagnetic duality on
ALE spaces. To this purpose let us consider a Maxwell
theory with a $\theta$-term \cite{ver, wittab}. The action for this theory is 
\be
S={1\over e^2}\int_X F\wedge *F - {i\theta\over 8\pi^2}\int_X F\wedge F  
\label{abelaction}
\ee
where $F=dA$ and $e$ is the electromagnetic coupling constant. 
The gauge field can be split into $A=A_{cl} + A_{qu}$ where, 
in order to have a globally defined gauge 
connection on $X$, $A_{cl}$ is 
a linear combination of the instanton potentials (\ref{monopot}) with
integer coefficients. As a consequence, the classical part of the  
field strength $F_{cl}$ will have in the basis (\ref{baschange}) the expansion
$F_{cl}=\sum_i 2\pi m^i c_1(\ct_i)$ with $m^i\in\zet$ given by 
\be
\int_{\Sigma_i} F_{cl} = 2\pi m^i 
\label{fexpansion}
\ee
The classical action takes the form
\be
S_{cl}[m^i] = {4\pi^2\over e^2} m^iG_{ij}m^j - {i\theta\over 2} m^iQ_{ij}m^j
= ({4\pi^2\over e^2} - {i\theta\over 2}) m^i (C^{-1})_{ij} m^j
\label{classaction}
\ee
because for ALE instantons $G_{ij}=Q_{ij}=(C^{-1})_{ij}$, as it follows
from the absence of anti-self-dual two-forms and (\ref{intertau}).
The partition function for this Maxwell theory is then given by
\be
Z_M = {{\rm det}^{\prime}(\Delta_F) \over {\rm det}^{\prime}
(\Delta_B)^{\mezzo} } 
\sum_{m^i\in\zzet}e^{-S_{cl}[m^i]}
\label{partsum}
\ee
where $\Delta_B$ and $\Delta_F$ are the kinetic operators
for the quadratic fluctuations (the only ones present in an abelian
theory) of boson and fermion fields respectively. As usual, the primes
mean that the determinants are to be computed in the functional space 
orthogonal to the zero-modes. In the case under consideration, 
this restriction 
is immaterial since there are neither fermionic nor bosonic 
neutral zero-modes,
as we shall see in section 5. Moreover,
specializing to a supersymmetric theory, the functional 
determinants exactly
cancel. 
Putting $\tau= {4\pi i\over e^2}+ {\theta\over 2\pi}$
the partition function becomes
\be
Z_M = \sum_{m^i} e^{i\pi\tau (m^iC^{-1}_{ij}m^j)} = \eta(\tau)^{N-1}
\sum_{\Lambda} \chi_{_\Lambda}(\tau) 
\label{partcart}
\ee
where $\eta(\tau)$ is the Dedekind function and 
$\chi_{_\Lambda}(\tau)$ are 
characters of the unitary representations of highest weight, 
$\Lambda$, of the affine Lie algebra associated to $\wt{\Delta}_\G$ 
at level one \cite{gin}.  
For $A$-type ALE spaces one has the affine Lie algebra of $SU(N)$ at level
one. The generalized electromagnetic duality (S-duality) is induced by the 
transformation \cite{ver, wittab}
\be
{\mbox {S}}: \quad \tau \to -{1\over\tau}
\label{sduality}
\ee
The characters $\chi_{_\Lambda}$ provide a unitary 
representation of S, in the
sense that
\be
\chi_{_\Lambda}(-{1\over \tau}) = {1\over \sqrt{N}} 
\exp(2\pi i \Lambda \Lambda^\prime)  \chi_{_{\Lambda^{\prime}}} (\tau)
\label{sdualchi}
\ee
From (\ref{sdualchi}) and $\eta(-1/\tau)= \sqrt{i\tau}\eta(\tau)$
it follows that, summing over $\Lambda$, the
S-transformed partition function is proportional to  $\chi_{_0}(\tau)$, the
character of the singlet representation of the affine Lie algebra $SU(N)$,
hence (\ref{partcart}) 
is not invariant under the transformation (\ref{sduality}). 
The lack of
S-invariance of a Maxwell theory on ALE spaces, unlike what 
happens on other \hk
manifolds \cite{ver} ({\it e.g.}~the four-torus $T4$ and the 
K\"ummer's third 
surface
$K3$) is due to the properties of the lattice of the charges $m^i$. 
For $T4$
and $K3$ the relevant lattices are $L_{(4,4)}$ and $L_{(3,19)}$  
respectively,
while for \mc ALE instantons one has the weight lattice of 
$SU(N)$. The former
are even, self-dual, Lorentzian lattices, while the latter 
is an
$(N-1)$-dimensional Euclidean lattice which is neither even 
nor self-dual.
Additional terms in the action which arise upon coupling 
to gravity
do not make the above result consistent with the 
expected modular
transformation of $Z$ for compact manifolds \cite{wittab}, 
which reads
\be
Z(-{1\over\tau}) = \tau^u \bar\tau^v Z(\tau) 
\label{modzeta}
\ee
where $u=(\chi_{_E}-\tau_{_H})/4$ and $v=(\chi_{_E}+\tau_{_H})/4$, 
$\chi_{_E}$
being the Euler characteristic and $\tau_{_H}$ the Hirzebr\"uch 
signature of
the manifold. For \mc ALE metrics $\chi_{_E}=N$ and 
$\tau_{_H}=1-N$, so that
$u=N/2-1/4$ and $v=1/4$, and only the dominant powers of $\tau$ in
(\ref{modzeta}) agree with the transformation properties of $Z_M$ 
under (\ref{sduality}). 
In view of this result, it appears that the issue of S-duality 
on ALE spaces 
must be more carefully reconsidered \cite{vw}.

\section{\KN Construction of \YM Instantons on ALE Spaces}
\setcounter{equation}{0}
\subsection{The ADHM Construction on $\real^4$}

Before embarking in the KN construction, we would like to recall
few relevant facts about the ADHM construction on $\real^4$ which will be 
carried over to ALE spaces.

Self-dual $SU(2)$ connections on $S^4$, can be put into one to one 
correspondence with holomorphic vector bundles of rank $2$ over  
$\complex P^3$
admitting a reduction of the structure group to its compact real form. 
The ADHM construction \cite{adhm} gives all 
these holomorphic bundles and consequently all $SU(2)$ connections 
on $S^4$.
The construction is purely algebraic and we find it more convenient 
to use, at
the beginning, quaternionic notations. The points, $x$, of the 
one-dimensional 
quaternionic space $\bb{H}\equiv \complex^2\equiv\R4$ can be 
conveniently
represented in the form $x=x^\m \s_\m$, with $\s_\o=\unita$ 
and $\s_r=i\s_r^P$. 
The conjugate of $x$ is $x^\dagger = x^\m \s_\m^\dagger$. 
A quaternion is said to be real if it is proportional to $\unita$ 
and imaginary if it has vanishing real part. 
The identity $\s_\m \s_\n^\dagger = \delta_{\mn}+i \eta_{\mn}^r \s^P_r$, 
where the $\eta_{\mn}^r$'s are the 't Hooft symbols,
can be used to write the one-instanton solution of Belavin, Polyakov, 
Schwarz and Tyupkin \cite{bpst} in the form
\be
A=\Im\biggl({x^\dagger dx\over |x|^2+\lambda^2}\biggr)
\label{treuno}
\ee
where $\Im(q)=(q-q^\dagger)/2$.
To find the number of moduli which parametrize this solution we act on 
it with the transformations of the symmetry group of the YM action,
which turns out to be the conformal group\footnote{In quaternionic notations 
the conformal group is $SL(2,\bb{H})$.}. 
We find in this way that only the five-parameter transformations
of the type 
\be
x\mapsto\cd=\mu(x_\o-x),\quad \mu\in\bb{R},\quad x_\o\in\bb{H}
\label{tredue}
\ee
deform the solution. The inclusion of the three parameters related to
global $SU(2)$ rotations in (\ref{tredue}) can be accomplished by
promoting $\mu$ to be a quaternion.

The structure of
(\ref{treuno}) and (\ref{tredue}) can be generalized to the case of
instantons with arbitrary winding number, $k$, by replacing
in (\ref{treuno}) $x$ with a column of $k+1$ quaternions,
$x\rightarrow U(x)\in\bb{H}^{k+1}$ and in (\ref{tredue}) $\m$ with 
a $k$-dimensional quaternionic vector, $\mu \rightarrow q\in\bb{H}^k$, and
at the same time $x_\o$ with a $k\times k$ matrix of quaternions, 
$x_\o \rightarrow a_\o$. By accomodating the vector $q$ and the 
matrix $a_\o$ in a single $(k+1)\times k$ matrix, $a= \pmatrix{a_\o\cr q}$, 
$\cd$ in 
(\ref{tredue}) can be generalized to $\cd=a+bx$, where $b$ is 
another $(k+1)\times k$ quaternionic matrix and $bx$ means multiplication 
of each element of $b$ by the quaternion $x$.

The (anti-hermitean) gauge connection is written in the form 
\be
A_\mu=U^\dagger\de_\mu U
\label{trecinque}
\ee
where $U$ is a $(k+1)\times 1$ matrix of quaternions providing an
orthonormal frame of $\Ker \cd^\dagger$. In formulae
\be
{\cal D}^\dagger U = 0
\label{trequattro}
\ee
\be
U^\dagger U =\unita_2
\label{trequattrobis}
\ee
where $\unita_2$ is the two-dimensional identity matrix. The 
constraint (\ref{trequattrobis}) ensures that $A_\mu$ is an element 
of the Lie algebra of the $SU(2)$ gauge group.
The condition of self-duality on the field strength of 
(\ref{trecinque}) is imposed by restricting the matrix $\cd$ to obey
\be
\cd^\dagger\cd=\Delta\otimes\unita_2
\label{realita}
\ee 
with $\Delta$ an invertible hermitean $k\times k$ matrix (of complex numbers). 
In addition to the gauge freedom (right multiplication by
a unitary quaternion) we have the freedom to left multiply
the matrices $a , b$ by a unitary $k\times k$ quaternionic matrix. 
These symmetries that can be used to simplify the expressions
of $a$, $b$ and $U$.

The extension of (\ref{treuno}) to the case of gauge groups other than 
$SU(2)$ is accomplished by further enlarging the dimensions of $\cd$, while 
equations (\ref{trecinque}), (\ref{trequattro}), (\ref{trequattrobis})  
and (\ref{realita}) remain formally unchanged.
 
For instance, for $Sp(2n)$\footnote{Using physicist's nomenclature we call 
$Sp(2n)$ the group of $2n$-dimensional matrices, $g$, with the 
property $g^T J g = J$, where $J$ is the symplectic (antisymmetric) metric.}
gauge groups $a$ and $b$ are $(k+n)\times k$ matrices of quaternions, 
where $b$ can always be cast in the standard form
\be
b=\pmatrix {\unita_{k\times k}\cr O_{n\times k}}
\label{boh}
\ee
by exploiting the simmetries of the construction, 
and $U$ becomes a $(k+n)\times n$ matrix of quaternions.

With minor changes a similar procedure can be used to construct $U(n)$
connections. In this case $U$ and $\cd$ can be taken as $(2k+n)\times n$ and
$(2k+n)\times 2k$ complex matrices, respectively. To impose the constraint
(\ref{realita}) it
is convenient to split $\cd$ into two matrices, $\cd_r=a_r+b_r x$, 
$r=1,2$, of $(2k+n)\times k$ elements each. 
The condition $\cd^\dagger\cd\propto \unita_2$ leads to
\be
a^\dagger_r a_s=\nu\delta_{rs} 
\label{tresei}
\ee
\be
b^\dagger_r b_s=\nu^\prime \delta_{rs},\quad 
\epsilon_{rs}b^\dagger_s a_t=\epsilon_{st}a^\dagger_s b_r
\label{treseibis}
\ee
where $\nu, \nu^\prime$ are $k\times k$ hermitean matrices and 
$\epsilon_{rs}$ is the two-dimensional anti-symmetric tensor.
Using (\ref{treseibis}) and the $U(k)$ symmetry of the construction, 
$a, b$ can be cast into form
\be
a=\pmatrix{A &-B^\dagger\cr B & A^\dagger\cr s & t^\dagger\cr},\qquad 
b=\pmatrix{\unita & 0\cr 0 &\unita\cr 0 & 0\cr}
\label{tresette}
\ee
where $A, B$ and $s, t^\dagger$  are $k\times k$ and $n\times k$ complex 
matrices 
respectively, subjected to the constraints coming from (\ref{tresei}).

For $U(n)$ gauge groups, the set 
\be
M=\{A,B;s,t  \, \vert \; A,B\in \End(V); s,t^\dagger\in \Hom(V,W)\}
\ee
where $V, W$ are  $k$-dimensional and $n$-dimensional complex 
vector spaces respectively, is often called a set of ADHM data. 
The real 
dimension of $M$ is $\dim(M)=4k^2+4kn$.

It is very much inspiring to reformulate the ADHM construction
in the language of the previous section. To this end we 
first notice that, in
terms of the matrices  $A, B, s, t^\dagger$, the condition for 
$\cd^\dagger\cd$
not to have off-diagonal blocks becomes
\be
[A,B]+ts=0
\label{treotto}
\ee
while the conditions for the diagonal blocks of $\cd^\dagger\cd$ to be
all proportional to the same $k\times k$ hermitean matrix, $\Delta$, is
\be
([A,A^\dagger]+[B,B^\dagger])-s^\dagger s+tt^\dagger =0
\label{trenove}
\ee
Observing that the three closed two-forms
\bea
\omega_\ccomplex &=&\Tr (dA\wedge dB)+\Tr (dt\wedge ds)\cr
\omega_\rreal &=&\Tr (dA\wedge dA^\dagger)+\Tr (dB\wedge dB^\dagger)-
\Tr (ds^\dagger\wedge ds) + \Tr (dt\wedge dt^\dagger)
\label{tredieci}
\eea
are invariant under the transformations $A\mapsto gAg^\dagger, s\mapsto
gsv^\dagger$ with $g\in U(k)$ and $v\in U(n)$ (and the same for 
$B, t^\dagger$), one concludes that the moment maps for the triholomorphic
$U(k)$ isometries are exactly given
by the l.h.s. of the ADHM 
equations (\ref{treotto}) and (\ref{trenove}). 
In the r.h.s. of these equations, there is no deformation parameter, 
such as the $\z$'s appearing in (\ref{duequattro}),
as a consequence of the fact that the base manifold, $\R4$, 
is undeformable. Equations (\ref{treotto}) and (\ref{trenove}) 
yield $3k^2$ real constraints on $M$. Taking into account the residual
$U(k)$ invariance, the \hkq of $M$ with respect to $U(k)$  
has dimension $4k^2+4kn-3k^2-k^2=4kn$, and coincides with the framed 
moduli space of $SU(n)$ self-dual connections on $\R4$.

\subsection{The ADHM construction on ALE Spaces}

The initial step in the KN construction (for $U(n)$ gauge groups)
is to give a set of ADHM data $M=\{A,B;s,t;\xi\}$ where $\xi\in \Xi$, 
$A$ and $B$ are $\G$-equivariant\footnote{$\G$-equivariance simply means 
that the matrices $\xi$, $A$ and $B$ are naturally decomposed into  
$n_i\times{n_j}$-dimensional blocks.} endomorphisms of a $k$-dimensional 
complex  vector space, $V$, and $s,
t^\dagger$ is a pair of homomorphisms between $V$ and  
an $n$-dimensional complex vector
space $W$. Both $V$ and $W$ are  $\G$-modules, 
\ie~they admit the decomposition  
$V=\oplus_i V_i\otimes R_i$, $W=\oplus_i W_i\otimes R_i$,  where
$V_i\sim\complex^{v_i}$, $W_i\sim\complex^{w_i}$. Therefore 
\be
k = \dim(V) = \sum_i
n_i v_i \qquad n = \dim(W) = \sum_i n_i w_i 
\label{dimvw}
\ee
where the lower case letters, 
$v_i,
w_i$, stand for the dimensions of the corresponding vector spaces. Out of 
these data the matrix 
\be
{\cal D}=({\cal A}\otimes \unita - \unita \otimes \xi) \oplus \Psi 
\otimes \unita
\label{defd}
\ee
is constructed. In (\ref{defd}) we have used the definitions
\be
{\cal A} =\pmatrix{A &-B^\dagger\cr B &A^\dagger\cr}
\in ({\rm T}^*\otimes {\End}(V))_\G=
\oplus_{i,j}A_{ij}\Hom(V_i,V_j)
\label{defaa}
\ee
and 
\be
\Psi = (s\quad t^\dagger)
\in \Hom(S^+\otimes V,W)_\G=\oplus_i \Hom(S^+\otimes V_i,W_i)
\label{defab}
\ee
We have arranged the matrices $A$ and $B$ into a
quaternion of matrices ${\cal A}$, similarly to what has been done in 
(\ref{duedieci}) to represent the points $\xi$ of the manifold $\Xi$, 
or in (\ref{tresette}). $S^+$ is isomorphic to the two-dimensional 
complex vector space, $\complex^2$, 
and may be conveniently thought as the space of right-handed spinors.

The $(2k+n)|\G|\times 2k|\G|$ matrix $\cd$ represents the linear map
\be
\cd:(S^+\otimes V\otimes\ct)\mapsto(Q\otimes V\otimes\ct)\oplus
(W\otimes\ct)
\label{linmap}
\ee
where $\ct$ is the tautological bundle and $Q$, defined after 
(\ref{duecinque}), is to be
identified with $S^-$, the space of left-handed spinors (the dual of $S^+$).
Remembering that the base manifold, $X_{\z}$, is a smooth resolution 
of $\complex^2/\G$, one is led to consider only the $\G$-invariant 
part of (\ref{linmap}), $\cd_{\G}$. 
When $\cd$ is restricted in this way, it becomes a $(2k+n)\times 2k$ matrix
which plays the same role as the matrix $\cd$ of the previous section:
more precisely in the matrix ${\cal A}\oplus\Psi$ we recognize the matrix 
$a$ of (\ref{tresette}). 

In order to generalize (\ref{trequattro})
we introduce the adjoint of the $\G$-restriction of (\ref{linmap}) as the 
mapping 
\be
\cd^\dagger_\G:{\cal V}\oplus {\cal W}\mapsto {\cal U}
\label{quattrouno}
\ee
where 
\bea
{\cal U}&\equiv& S^+\otimes (\bar V\otimes\ct)_{\G} \cr 
{\cal V}&\equiv& (\bar Q\otimes \bar V \otimes\ct)_{\G}\cr 
{\cal W}&\equiv& ({\bar W}\otimes\ct)_{\G}
\label{decuvw}
\eea
and $\bar Q, \bar V, \bar W$ denote the trivial (\ie~product) vector
bundles over $X_\zeta$ with fiber $Q, V, W$ 
respectively\footnote{In view of
the following applications, we refer more
rigorously in (\ref{decuvw}) to bundles, and not to vector spaces, because
we will have to consider sections on them, \ie~
locally defined functions of the variable $x\in X$.}.

Once $\xi\in\Xi$ has been restricted to lie on the ALE space $X$
by imposing 
\bea
[\alpha,\beta] &=&\zeta_{\ccomplex}\cr
[\alpha,\alpha^\dagger]+[\beta,\beta^\dagger] &=&
\zeta_{\rreal}
\label{kroeq}
\eea
self-duality of the resulting YM connection follows from 
the condition $\cd^\dagger_{\G}\cd_{\G}=\Delta\otimes\unita$ with 
$\Delta$ a hermitean $k\times k$ matrix and $\unita$ the $2\times 2$ identity 
matrix in $S^+$. This is accomplished by imposing the deformed
version of the ADHM equations (\ref{treotto}) and (\ref{trenove})
\bea
[A,B]+ts &=&-\zeta_{\ccomplex}\cr
[A,A^\dagger]+[B,B^\dagger]-s^\dagger s+tt^\dagger &=&-
\zeta_{\rreal}
\label{adhmeq}
\eea
where $\z=\oplus_{i=0}^{r-1} \z_i \unita_{v_i}
\in \real^3\otimes{\cal Z}^*_{_V}$, with 
${\cal Z}^*_{_V}$ the dual to the center of the Lie algebra of 
$G_{_V}=\otimes_i U(v_i)$ and the 
parameters $\z_i$ in the r.h.s of (\ref{adhmeq}) are identified with 
those in the r.h.s. of (\ref{kroeq}), thanks to the homomorphism of 
$U(n_i)$ in $U(v_i)$.  

The condition (\ref{trequattro}) allows to identify the instanton bundle 
${\cal E}$ with
\be
{\cal E}={\Ker}\cd^\dagger_{\G}\subset {\cal V}\oplus{\cal W}
\label{kerbundle}
\ee
${\cal E}$ is then a complex vector bundle on $X_{\z}$, 
with typical fiber $W$ and rank $n=\dim(W)$.
The YM connection on ${\cal E}$ is given by 
\be
A_\mu=U^\dagger\nabla_\mu U
\label{connec}
\ee
where $U$ represent an orthonormal frame of sections of 
${\Ker}\cd_{\G}^\dagger$, \ie~a $(2k+n)\times n$ complex matrix obeying 
$\cd_{\G}^\dagger U=0$ and $U^\dagger U = \unita_n$.
Since $\bar Q, \bar V, \bar W$ are flat bundles,
the covariant derivative on $({\bar Q}\otimes{\bar V}\otimes\ct)_{\G}$
takes the form $\nabla_\mu=(\partial_\mu+A_\mu^{\ct})$, with 
$A_\mu^{\ct}$
the (self-dual) connection on the tautological bundle $\ct$.
For \mc ALE metrics the abelian connections on $\ct_i$ are given in 
(\ref{monopot}).

\subsection{Particular Solutions on ALE Spaces and 
$\G$-invariant Instantons}

Not unlike what happens on $\real^4$, 
the generalized ADHM equations on ALE spaces 
can be solved explicitly only in the simplest cases.

On $\R4$ for the group $SU(2)$ solutions with generic winding number 
$k$, but with a number of 
parameters (moduli) smaller than the dimension of the moduli space have been
found. For example the  well-known 't Hooft ansatz \cite{acto}
$A^r_\m = -\bar\eta^r_{\mn} \de^\n \log \phi$, where
\be
\phi_{_{t'H}} = 1 + \sum_{i=1}^k {\rho_i^2 \over (x-x_i)^2}
\label{hooft}
\ee
contains only $5k$ parameters, the centers and sizes of the instantons, 
instead of the expected $8k-3$.
Similarly the conformally invariant solution of Jackiw, Nohl and Rebbi (JNR) 
\cite{jnr} with
\be
\phi_{_{JNR}} = \sum_{i=1}^{k+1} {\l_i^2 \over (y-y_i)^2}
\label{jnrsolution}
\ee
effectively depends only on $5k+4$ parameters, without a clear geometrical
meaning. The expected number of moduli is matched by (\ref{hooft}) 
for $k=1$ and by (\ref{jnrsolution}) 
for $k=1,2$. Actually also for $k=3$ one can find self-dual
connections with the right number of moduli by directly solving 
the ADHM equations \cite{cws}.

It is not difficult to relate (\ref{hooft}) and 
(\ref{jnrsolution}) to $SU(2)$ instantons
on orbifolds $X_{\o}=\R4/\G$, whose smooth resolutions are precisely the ALE 
manifolds we are interested in. 
In the orbifold limit, one is lead to consider 
\sd connections on $\real^4$ satisfying $A_\m(\g x) = A^{\W_{\g}}_\m(x)$
for $\gamma\in\G$, \ie~invariant under $\G$ up to gauge transformations. 
The condition of strict $\G$-invariance, $\W_{\g} = \unita, \forall\g\in\G$, 
is particularly simple to achieve for $\G=\zet_2$, \ie~for the 
orbifold limit of EH instanton. 
One simply has to place the centers appearing in (\ref{hooft}) or 
(\ref{jnrsolution}) in
$\zet_2$ symmetric configurations. In particular for $k$ even one may take
\be
\phi = 
1+\sum_{i=1}^{k/2} \left({\rho_i^2 \over (x-x_i)^2}+{\rho_i^2 
\over (x+x_i)^2}\right)
\label{tequiv} 
\ee
and for $k$ odd  
\be
\phi = \sum_{i=1}^{{(k+1) /2}} \left({\l_i^2 \over (y-y_i)^2}+
{\l_i^2 \over (y+y_i)^2}\right)
\label{jnrequiv} 
\ee
The number of free parameters in (\ref{tequiv}) and 
(\ref{jnrequiv}) is clearly reduced 
with respect to (\ref{hooft}) and 
(\ref{jnrsolution}) and it happens to coincide with 
the actual dimensions of the moduli space of the instanton connections 
only for bundles with second Chern class $\kappa = 1/2, 1$ and $3/2$, 
corresponding to $k=1,2$ and $3$ in the above formulae.
The ansatze (\ref{tequiv}) and (\ref{jnrequiv}) can be generalized to the 
other
\mc ALE metrics, where they amount to place the centers in a $\zet_N$
symmetric way. 

When $X$ is a generic smooth ALE manifold, another possibility 
is to consider restricted classes of solutions of the
self-duality equations. One starts from the very 
simple and inspiring ansatz valid for $SU(2)$ given in 
\cite{bcc} for the case of \mc ALE metrics (\ref{multicenter} which reads
\be
A_\o = \mezzo A_\o^r\s_r^P = \mezzo \vec{G}\cdot\vec\s^P \quad\quad 
\vec{A} = \mezzo (\vec\omega (\vec{G}\cdot\vec\s^P) - 
V (\vec{G}\times\vec\s^P) )
\label{bccansatz}
\ee
where $V$ is defined in (\ref{potv}) and $\vec G$ is taken to be 
independent from the cyclic coordinate $t$. The general
($t$-independent) solution of the self-duality equations is given by 
\cite{bcc}
\be
\vec{G}= - V^{-1} \vec\nabla \log(H)
\label{bccsolution}
\ee
with $H=\sum_{i=1}^n {\lambda_i /|x-x_i|}$. Regularity of the gauge action
restricts the sum in $H$ to a subset of the $N$ 
centers, $\vec x_i$, appearing in $V$ ($n\leq N$). The 
resulting $SU(2)$ connections
have second Chern class $\kappa=n-1/N \leq N-1/N$. For $n=N$,
the bound is saturated, 
$\vec{G}$ equals $- V^{-2} \vec\nabla V$ and the $SU(2)$ gauge connection 
(\ref{bccansatz}) 
coincides with the self-dual spin connection on the tangent bundle to $X$ 
\cite{bcc}. 

The ansatz (\ref{bccansatz}) may be cast into a form which more clearly 
resembles (\ref{hooft}) and (\ref{jnrsolution}).
In fact, introducing the inverse tetrad $E_a$ with 
\be
E_\o = E_\o^\m\de_\m = V^{\mezzo} {\de\over\de t} \quad\quad
E_l = E_l^\m\de_\m = V^{-\mezzo} ({\de\over\de x^l} - \omega_l 
{\de\over\de t})
\label{inversetet}
\ee 
one finds that (\ref{bccansatz}) can be generalized to 
\be
A^r_\m dx^\m = -\bar\eta^r_{ab} e^a E^b(\log H)
\label{bccth}
\ee 
with $H$ now not necessarily independent from the coordinate $t$.
Substituting (\ref{bccth}) into the self-duality equation $F=* F$, 
one gets for $H$
\be
{\nabla_a\nabla^a H\over H}=0
\label{covlaplace}
\ee
where $\nabla_a$ is the covariant derivative on $X$. To derive 
(\ref{covlaplace}) the three covariantly constant complex structures,
$I,J,K$, introduced in  subsection 2.1, have been identified 
with $\bar\eta^i_{ab}$,
and we have used the equations $E_a H=E_a^\mu\de_\mu
H=\nabla_aH$ and $\nabla_a\nabla_bH=\nabla_b\nabla_aH$. The solutions of
(\ref{covlaplace}) 
can be written in the form $H(x)=H_\o + \sum_i^n G(x,x_i)$, where 
$H_\o$ is a constant and $G(x,x^\prime)$ is the scalar  propagator 
in the background of the \mc metric. $G(x,x^\prime)$ has been explicitly 
computed by Page
\cite{pag}. Depending on whether the constant $H_\o$ is zero or not, 
(\ref{bccth}) 
turns out to have second Chern class $\kappa=n-1/N$ or $\kappa=n$,
respectively. The ansatz (\ref{bccth}) 
seems to be more flexible than the fullfledged
solutions of the KN-ADHM equations as far as the computation of zero-modes
is concerned. Moreover, the ansatz (\ref{bccth}) can be used 
to generate self-dual $SU(2)$ connections on other four-dimensional
\hk manifolds, such as, for instance, the $K3$ surface.

\section{Topological Properties of \YM Instanton Bundles on ALE Manifolds}
\setcounter{equation}{0}
\subsection{General Setting}

In this subsection we illustrate how to compute some topological
invariants of the instanton bundle ${\cal E}$, such as the first and second 
Chern class and the dimension of its moduli space.
Recalling
\bea
V &=&\oplus_i V_i\otimes R_i\cr
W &=&\oplus_i W_i\otimes R_i\cr 
\ct &=&\oplus_i \ct_i\otimes R_i
\label{quattrotrebis}
\eea
we can write
\bea
{\cal U} &=& S^+\otimes(\bar V\otimes\ct)_\G = 
S^+\otimes(\oplus_i\bar V_i\otimes\ct_i)\cr
{\cal W} &=& (\bar W\otimes\ct)_\G = \oplus_i \bar W_i\otimes\ct_i\cr
{\cal V} &=& (\bar Q\otimes \bar V\otimes\ct)_\G = \oplus_{i,k} A_{ik} \bar
V_i\otimes\ct_k 
\label{quattrodue}
\eea
where we have used the decompositions (\ref{duesei}), (\ref{duesette}).

From these formulae the first and second Chern class of 
${\cal E}$ can be expressed as \cite{kn}
\bea
c_1({\cal E}) &=&c_1({\cal V})+c_1({\cal W})-c_1({\cal U})=
\sum_i (w_i-\sum_j\tilde C_{ij}v_j)c_1({\cal T}_i) =
 \sum_i u_i c_1({\cal T}_i) \cr
c_2({\cal E}) &=& \sum_i u_i c_2({\cal T}_i) + {k\over |\G|}
\label{chern}
\eea
with
\be
u_i \equiv w_i-\sum_j\tilde C_{ij}v_j \qquad i=1,\cdots,r-1
\label{quattrotre}
\ee
In equations (\ref{chern}) and (\ref{quattrodue}) the term with $i=0$ does 
not actually appears in the sums over $i$, because the triviality of 
${\cal T}_0$ implies $c_1({\cal T}_0)=0$ and $c_2({\cal T}_0)=0$. 
Making use of the relation 
$\sum_{j=\scriptscriptstyle 0}^{r-1}\tilde C_{ij}n^j=0$ 
and
recalling that $n_{\o}=1$, it is possible to consistently extend the
definition (\ref{quattrotre}) to $i=0$, by putting   
\be 
u_\o=w_\o+\sum_{i\neq \scriptscriptstyle 0} n_i(w_i-u_i) 
\label{quattroquattro}
\ee
If the structure group
of the bundle ${\cal E}$ is restricted to $SU(n)$, then $c_1({\cal E})=0$ 
and 
one immediately finds $u_i=0$ for $i=1,\cdots,r-1$.

The rank of the instanton bundle, ${\cal E}$, defined by (\ref{quattrouno}) 
and (\ref{kerbundle}) can be computed from the formula
\be
{\mbox{rank}}{\cal E}=\dim({\cal E}) = \dim(\Ker(\cd^\dagger)) =
\dim({\cal V})+\dim({\cal W})-\dim({\cal U})
\label{ranke}
\ee
Using $A_{ij} = 2 \d_{ij} - \tilde C_{ij}$, one finds as expected
\be
\dim({\cal E})=\sum_i w_in_i=n
\label{quattroduebis}
\ee
with the integers $w_i$ being the dimensions of the subspaces $W_i$.

Recalling the general discussion after (\ref{tredieci}), it can be shown 
that the framed moduli space of connections on ${\cal E}$,
${\cal M}_{\cal E}$, is itself a \hk manifold. 
Indeed ${\cal M}_{\cal E}$ is given by the equivalence class of data 
$({\cal A}, \Psi)\in M$ 
satisfying the ADHM equations (\ref{adhmeq}), up to transformations of 
$G_{_V}=\otimes_i U(v_i)$. 
From the \hk quotient construction it follows that 
${\cal M}_{\cal E}={\cal P}_{\cal E}/G_V$, with ${\cal P}_{\cal E}$
a $G_V$ principal bundle on ${\cal M}_{\cal E}$, analogous to $P_\z$ 
in (\ref{duequattro}).
From (\ref{defaa}) and (\ref{defab}), the real dimension
of ${\cal M}_{\cal E}$ is computed to be
\bea
\dim({\cal M}_{\cal E}) &=& \dim(M) - 4\dim(G_{_V})= 
\dim({\cal P}_{\cal E}) - 4\dim(G_{_V})\cr
&=& 2\sum_i 
(\sum_j A_{ij}v_iv_j+4 v_iw_i-4v_i^2)=2\sum_i v_i(u_i+w_i)
\label{purepeggio}
\eea 
To clarify the ADHM construction in the general case let us first 
discuss few explicit examples  which will be also useful for the 
applications presented in section~6. We start by
noticing that, in order to get a finite gauge action, the instanton 
connection
must be asymptotically equal to a pure gauge. However, the gravitational
instanton background allows for the existence of non-trivial holonomies
at infinity. This means that, if one parallel transports the typical fiber 
of the instanton  bundle, ${\cal E}$, around infinity, it will be 
transformed according
to  some representation of $\G$. This possibility is the main
novelty in the KN construction on ALE spaces with respect to the 
standard ADHM construction on $\R4$.

As a first example let us take $W=R_i$ and $V=\emptyset$. In this case the
instanton bundle ${\cal E}$  coincides with the elementary bundle $\ct_i$ 
(${\cal E}=\ct_i$), 
so that, recalling the decomposition formulae (\ref{quattrotrebis}) 
and equations (\ref{quattrotre}) and (\ref{quattroquattro}), 
one gets for the dimension vectors, $v=0$ and $u=w=(0,\ldots,1,\ldots,0)$, 
with the 1 in the $i$-th position. It follows from
(\ref{purepeggio}) that the moduli space of ${\cal E}$ is 0-dimensional,
\ie~the connection on the elementary  bundle, $\ct_i$, has no free 
continuous parameter, as explicitly found in 
(\ref{monopot}) for the case of \mc ALE metrics. Strictly speaking the 
tautological bundles $\ct_i$ correspond to singular limits of the KN 
construction in which $A,B,s,t$ are absent, while at the same time 
the ADHM equations (\ref{adhmeq}) 
loose their force because the instanton connection is abelian.

The simplest non-abelian instanton bundle is the
$SU(2)$ bundle associated to the natural (two-dimensional) homomorphism, 
$\rho_{_Q}(\gamma)$, of $\G$ in $SU(2)$.
For $\G=\zet_N$, the two-dimensional representation space decomposes
according to $Q=R_1\oplus \bar R_1$. 
The corresponding decomposition of $W$ is associated with the dimension 
vector
$w=(0,1,0,\ldots,0,1)$. The homomorphism $\rho_{_Q}(\gamma)$ of $\zet_N$ 
in $SU(2)$ implies that, if we parallel transport around infinity 
$m$ times a section $\phi$ of ${\cal E}$, it transforms according to
\be
\phi= \pmatrix{\phi_1\cr\phi_2\cr} 
\mapsto \Tr P \exp (i\oint Adx) 
\pmatrix{\phi_1\cr\phi_2\cr} =
\pmatrix{e^{{2\pi i\over N}m} & 0 \cr 0 & e^{-{2\pi i\over N}m}\cr}
\pmatrix{\phi_1\cr\phi_2\cr}
\label{parallel}
\ee
where $A$ is the connection on ${\cal E}$ and $P$ denotes path-ordering. 
In order to identify 
${\cal E}$, one may recall that for $SU(2)$ bundles
$c_1({\cal E})=0$. In view of (\ref{chern}) this condition 
is equivalent to $u_i=0$ for $i\neq 0$. To determine 
$u_{_0}$ 
and the vector $v$ we solve the system of $N-1$ homogeneous
equations (\ref{quattrotre}). This leads to
$v=(k_1-1,k_1,\ldots,k_1)$ and $u=(2,0,\ldots,0)$, with $k_1$ a positive 
integer. The different choices of 
$k_1$ correspond to different instanton numbers, because 
$c_2({\cal E})=\kappa=k_1-{1\over N}$. Using (\ref{purepeggio}), 
we find the dimension of the moduli space to be $\dim({\cal M}_{\cal E})
= 8\kappa+{8\over N}-4=8k_1-4$. For $k_1=1,\ldots,N$ these solutions 
are topologically equivalent to those in \cite{bcc}, generalized here 
by the ansatz (\ref{bccth}), (\ref{covlaplace}).

In the next section, we will devote particular attention to 
rank-two ($n=2$) $SU(2)$ bundles on the EH instanton.
In this case $\G=\zet_2$, so that $R_1=\bar R_1$ and (\ref{quattroduebis}) 
is satisfied by 
the three choices $w=(0,2)$, $w=(2,0)$ and $w=(1,1)$. 
The last case will not be considered here because it corresponds to a
non-vanishing first Chern class. In the other two cases one has $u=(2,0)$.
For $w=(0,2)$, using (\ref{quattrotre}) one finds $v=(k_1-1,k_1)$ and 
(\ref{purepeggio}) gives
\be
\dim({\cal M}_{\cal E})=8k_1 - 4
\label{quattrosei}
\ee
The lowest $k_1$ values, $k_1=1$ and $k_1=2$, correspond to an instanton 
moduli
space of dimensions 4 and 12 respectively.
The non integral value of the corresponding second Chern classes,
$\kappa={1/2}$ and $\kappa={3/2}$, is a consequence of the
non-trivial holonomy of the connection at infinity, 
where the instanton bundle ${\cal E}$ coincides with $\ct_1\oplus\bar\ct_1$.
$\bar\ct_1$ is the bundle conjugate to $\ct_1$, \ie~the one with 
monopole charge opposite to $\ct_1$, as shown in (\ref{parallel}). 

For $w=(2,0)$, one finds $v=(k_2,k_2)$ and $\dim({\cal M}_{\cal E})=8k_2$.
These $SU(2)$ bundles descend from $SU(2)$ bundles on $\R4$ with 
even instanton number $k=2k_2$. Indeed, the connection has trivial 
holonomy at infinity, \ie~it coincides with the trivial connection on 
the bundle $\ct_\o\oplus\ct_\o$.

\subsection{Index Theorems}

Generalizing the inverse construction of Corrigan and Goddard \cite{cg}, 
\Kro and \Nak have also carried over to ALE spaces
the difficult part of the ADHM construction, \ie~they have shown the 
uniqueness and completeness of the construction \cite{kn}.
Using hard analysis on ALE spaces (mainly Sobolev spaces 
with properly defined norms) \Kro and \Nak have shown that 
the ADHM data $W_i, V_i$ are related to the space of bounded harmonic
scalars, $W_i={\cal H}(\Delta, {\cal E}\otimes\ct_i)$, and to the space of
\zms of the Dirac operator, $V_i={\Ker}(\fey, {\cal E}\otimes\ct_i)$
\cite{kn}. Apart from checking the above identifications in some explicit 
examples, we are not able to reproduce their results in any easy way so that
we refer the reader interested in this aspect of the KN construction 
to the original paper \cite{kn}. We simply collect some relevant facts which
may be helpful in the computation of instanton effects on ALE spaces.
 
No neutral spinor zero-modes are expected on ALE spaces
since the index of the Dirac operator (for gauge singlets) is zero 
\cite{egh}.
However the presence of charged spinor zero-modes is guaranteed by 
a non-vanishing index of the Dirac operator coupled to the gauge bundle 
${\cal E}$. 
No Dirac \zm of wrong chirality is expected as well as no (normalizable) 
\zm of
the scalar Laplacian.

These results are obtained computing the classical topological invariants on 
ALE spaces. Standard formulae, 
which can be found in \cite{gpr}, lead for the
Euler characteristic, $\chi_{_E}$, and the Hirzebruch signature, 
$\tau_{_H}$, to the results 
\bea
\chi_{_E} &=&-{1\over 16\pi^2}\int\tr R\wedge *R+{1\over |\G|} = r\cr
\tau_{_H} &=&-{1\over 24\pi^2}\int\tr R\wedge R+\xi_s = r - 1 
\label{hirzeuler}
\eea
where $r$ is the rank of $\G$. The number of gauge singlet spin
${1/2}$ and  ${3/2}$ zero modes are given by
\bea
I_{1\over 2} &=&{1\over 192\pi^2}\int\tr R\wedge R+\xi_{1\over 2}=
-{\tau_{_H}-\xi_s\over 8}+\xi_{1\over 2}\cr
I_{3\over 2} &=&-{21\over 192\pi^2}\int\tr R\wedge R+\xi_{3\over 2}=
{21\over 8}
(\tau_{_H}-\xi_s)+\xi_{3\over 2}
\label{quattrosette}
\eea
The so called $G$-index theorems allow the 
computation of the boundary corrections, $\xi$'s, for which one finds 
\cite{gpr} 
\bea
\xi_{1\over 2} &=&
{1\over |\G|}\sum_{\gamma\neq e}{1\over 2-\chi_{_Q}(\gamma)}\cr
\xi_s &=& 4\xi_{1\over 2}-1+{1\over |\G|}\cr
\xi_{3\over 2} &=& 3\xi_{1\over 2}-2+{2\over |\G|}
\label{quattrootto}
\eea
where $\chi_{_Q}$ is the character of the representation 
$\rho_{_Q}$ of $\G$
and the sum runs over all 
elements of $\G$ different from the identity.

For \mc ALE spaces, where $\G=\zet_N$, the boundary corrections
can be explicitly computed. Recalling that for $\G=\zet_N$ any 
element $\gamma\in\zet_N$ satisfies $\gamma^N=1$, one gets
$\chi_{_Q}(\gamma)=\tr [{\mbox{diag}}(\epsilon^{m\over 2},
\bar\epsilon^{m\over 2})]=2\cos({2\pi m\over N})$ with 
$\epsilon=\exp ({4\pi i\over N})$ and $m=1,\ldots,N-1$.
From this result one finds 
\bea
\xi_{1\over 2}&=& {1\over 4N}\sum_{m=1}^{N-1}{1\over\sin^2
\left({\pi m\over N}\right)}={N^2-1\over 12N}\cr
\xi_s &=& {(N-1)(N-2)\over 3N}
\label{quattronove}
\eea
We remark that $\xi_s$ is zero for flat space ($N=1$) and for the EH
gravitational instanton ($N=2$).

The generalization of the index theorem for the spin ${1/2}$ complex to 
arbitrary representations, $T$, of the $SU(2)$ gauge group gives
\be
I_{1\over 2}^{(2j+1)}=\dim T \biggl(-{\tau+1\over 8}+{1\over 2}
\xi_{1\over 2}+{1\over |\G|}\biggr)+{\tr_T(T^aT^b)\over
\tr_Q(T^aT^b)}c_2({\cal E})+
\xi_{1\over 2}^{(T)}
\label{qdieci}
\ee
where $\dim T=(2j+1)$ and the second term on the r.h.s. is the gauge bulk 
contribution. The boundary correction for $\G=\zet_N$ is given by
\be
\xi_{1\over 2}^{(T)}={1\over |\G|}\sum_{\gamma\neq e}{\chi_{_T}(\gamma)
\over 2-\chi_{_Q}(\gamma)}={1\over N}\sum_{m=1}^{N-1}{\sin(2j+1)
{2\pi\over N}\over
2\sin {2\pi m\over N}(1-\cos {2\pi m\over N})}
\label{qundici}
\ee
In analogy with the case of the fundamental two-dimensional representation, 
one can compute $\chi_{_T}(\gamma)$ through the formula 
\be
\chi_{_T}(\gamma)= \tr[\diag
(\epsilon^{mj},\ldots,\epsilon^{-mj})]={\sin\left[(2j+1) {2\pi m\over
N}\right]\over\sin({2\pi m\over N})}
\label{chit}
\ee
Using the  notation of the previous section, we finally find 
\bea
I_{1\over 2}^{(2)}&=& k_1-1=v_0=\dim(V_0)\cr
I_{1\over 2}^{(3)} &=& 4k_1 -2 ={1\over 2}\dim({\cal M}_{\cal E})
\label{qdodici}
\eea
These results generalize those obtained in \cite{bfmr}.
Comparing the number of spin ${1/2}$ zero-modes in the adjoint 
representation of $SU(2)$ obtained there with (\ref{qdodici}), 
we find that the solutions of the heterotic string equations of 
motion in the background of the EH gravitational instanton, fulfilling the 
standard embedding condition, must have instanton number $\kappa={3/2}$.
For a generic ALE instanton, the $SU(2)$ connection corresponding
to the standard embedding is obtained taking $\kappa=|\G| - 1/|\G|$,
\ie~$k_1=|\G|$, from which $I_{1\over 2}^{(3)} =4|\G|-2$ follows.

\section{$SU(2)$ Gauge Instantons on the EH Manifold}
\setcounter{equation}{0}
We now specialize the discussion to $SU(2)$ instantons on the EH background 
expanding the results obtained in \cite{abfmr}.
In this case $\G=\zet_2$, the flat \hk manifold $\Xi$ is $\real^8$ and
the  \hk quotient is taken with respect to the group $G=U(1)$, 
since
the two irreducible representations of $\zet_2$ are one-dimensional 
($n_{\scriptscriptstyle 0} = n_1 = 1$).
The metric on the EH instanton \cite{eh} is
\be
ds^2_X={(dr)^2\over\unoar}+{r^2\over 4}(\sigma_x^2+\sigma_y^2)+
{r^2\over 4}\bigl(\unoar\bigr)\sigma_z^2
\label{ehmetric}
\ee
where the $\sigma_i$'s are the left-invariant one-forms of $SU(2)$. The 
formula
(\ref{ehmetric}) 
is explicitly derived in the appendix, using the hyper-K\"ahler quotient 
construction.
For future use we record here also the expression of the 
$U(1)$ instanton connection on the EH manifold, computed in \cite{eh}
\be
A^\ct =-{a^2\over r^2}(d\psi+\cos\theta d\phi)
\label{ultima}
\ee
(\ref{ultima}) is obtained from (\ref{monopot}), 
by changing to the variables used in (\ref{ehmetric}).

We already observed that the condition $c_1({\cal E})=0$ 
for rank two bundles can be satisfied in two different ways, with 
$u=(2,0)$
in both cases. For bundles which
descend from those with odd Chern class on $\R4$ one has $w=(0,2)$ and 
$v=(k_1-1,k_1)$, while
for bundles with even Chern class on $\R4$ one finds $w=(2,0)$ and 
$v=(k_2,k_2)$, where $k_1, k_2$ are positive integers. 

In the following we will see that for $k_1=1,2$ and for $k_2=1$
the KN ADHM equations may be solved explicitly, showing in particular 
that in the limit $a\to 0$ (\ie~in the orbifold limit $X_{\z} 
\rightarrow X_{\o}=\R4/{\zet_2}$) the solutions are invariant 
under $x\rightarrow -x$, \ie~are $\zet_2$-invariant instantons 
on $\R4$.

\subsection{$SU(2)$ Gauge Bundle with $c_2({\cal E})=\mezzo$}

This case was called the minimal $SU(2)$ instanton bundle 
in \cite{abfmr} because of its 
topological numbers $c_1({\cal E})=0, c_2({\cal E})=1/2$. 
Thanks to the simple form of the resulting gauge connection,
instanton dominated correlators around this background
could be explicitly computed in \cite{bbfmr}.
As we said
in the previous section, the vector spaces $V, W$ 
admit a decomposition as in (\ref{quattrotrebis}), 
with $w=(0,2), v=(0,1), u=(2,0)$.
In this case the matrices $A$ and $B$, appearing in (\ref{defaa})
are simply
absent. The map $\Psi$ is represented by a pair of $1\times 2$ 
complex matrices, \ie~by two two-dimensional complex vectors.
From (\ref{quattrouno}) we see that $\cd_\G^\dagger$ acts on the frame 
of sections, $U$, of the bundle ${\cal V}\oplus{\cal W}$ admitting the 
decomposition (\ref{quattrodue}). Since $A_{ij}$
off diagonal, we notice from the last formula in (\ref{quattrodue}) that, 
the sections of the space $\bar{V}_1$ are coupled only to
those of the trivial bundle, $\ct_0$. Moreover, the sections of 
$\ct$ are appropriately cast into the form of doublets, as 
in this way they are naturally 
acted upon by the elements of $\xi$, which are two by two matrices. 
Putting these observations together, we may write the complex
sections $u_{_U}$ appearing in $U$ in the form
\be
u_{_U}=\pmatrix{\vv_1\otimes\pmatrix{\phi_1\cr 0}\cr \vv_2\otimes
\pmatrix{\phi_2\cr 0}\cr\pmatrix{\w_1\cr \w_2}\otimes\pmatrix{0\cr \phi}}
\label{cinqueuno}
\ee
where $\vv_1, \vv_2 \in {\bar V}_1$; $\phi \in\ct_1$; 
$\phi_1, \phi_2 \in\ct_0$; $\w_1,\w_2\in {\bar W}_1$.
Following the arguments spelled out in the appendix, we see that 
$\G$-invariance restricts the matrices $\a$ and $\b$ in (\ref{duedieci}) 
to the form
\be
\alpha=\pmatrix{0&x_1\cr y_1&0\cr}\quad 
\beta=\pmatrix{0&x_2\cr y_2&0\cr}
\label{questaserve}
\ee
where $x_1$, $x_2$, $y_1$, $y_2$ are four complex coordinates on 
$\Xi=\real^8$.
The map $\Psi$ in (\ref{defab}) can be represented as
\be
\Psi = (s\quad t^\dagger)=\pmatrix{s_1&-\bar t_2\cr 
s_2&\bar t_1\cr}
\label{defpsi}
\ee
where $s_1$, $s_2$, $t_1$, $t_2$ are four complex parameters.

Combining (\ref{questaserve}) with (\ref{defpsi}), one can construct 
the matrix $\cd^\dagger_{\G}$
\be
\cd^\dagger_{\G} = \pmatrix{\a^\dagger&\b^\dagger&s^\dagger\cr
-\b&\a&t}
\label{cddagger}
\ee
Acting with (\ref{cddagger}) 
on (\ref{cinqueuno}), it is easily seen that the non trivial 
equations coming 
from the condition $\cd^\dagger_{\G}u_{_U}=0$, can be obtained from the 
reduced form of $\cd^\dagger_\G$ 
\be
\cd^\dagger_{\G} = \pmatrix{\xbar_1 &\xbar_2 &\bar s_1&\bar s_2\cr 
-y_2&y_1&-t_2&t_1\cr}
\label{cdrestricted}
\ee
acting on a similarly reduced form of $u_{_U}$.

Putting $\zeta_{\rreal}=-a^2$ and $\zeta_{\ccomplex}=0$ (see the 
discussion at the end of section 2.2) and introducing the definitions 
$x^2=|x_1|^2+|x_2|^2$ and $s^2=|s_1|^2+|s_2|^2$, the 
ADHM equations (\ref{adhmeq}), are solved by 
$y_1=\lambda x_1, y_2=\lambda x_2, t_1=\mu s_1, t_2=\mu s_2$ 
with $\lambda^2=1+{a^2/ x^2}, \quad\mu^2=1-{a^2/s^2}$.
By exploiting the $U(1)$ isometries of the
principal bundles $P_{\z}$ over $X_{\z}$ and ${\cal P}_{\cal E}$ over
${\cal M}_{\cal E}$, $\lambda$ and $\m$ can be chosen to be real. 
The orthonormal frame $U$ for the two-dimensional space 
${\Ker}\cd_{\G}^\dagger$ can be arranged in the $2 \times 4$ matrix
\be
U={1\over xs\sqrt{x^2+s^2}}\pmatrix{s^2 x_1&-\mu s^2\bar x_2\cr
s^2 x_2&\mu s^2\bar x_1\cr -x^2 s_1&\lambda x^2\bar s_2\cr
-x^2 s_2&-\lambda x^2\bar s_1\cr}
\label{ugauge}
\ee
where $x=\sqrt{x^2}$ and  $s=\sqrt{s^2}$.
In this setting $SU(2)$ gauge transformations correspond to 
$U \to U\Omega$ with $\Omega \in SU(2)$.

The last ingredient needed to compute the self-dual connection (\ref{connec}) 
is the abelian ($G=U(1)$) connection on the \tb $\ct$. Having reduced 
the frame $U$ to a pair of four-dimensional vectors, as explained before, 
we observe that the first two components of these vectors correspond to 
sections of $\ct_0$ which is trivial by construction. 
The covariant derivative, $\nabla_\mu$, appearing in (\ref{connec}) can 
thus be written as a $4 \times 4$ diagonal matrix of the kind 
$\nabla_\mu=\diag (\de_\mu,\de_\mu,\de_\mu+iA_\mu^\ct,\de_\m+iA_\mu^\ct)$. 
Putting $x_i=|x_i|e^{i\a_i}$ (see the appendix), the connection one-form on 
$\ct$ becomes
\be
A_\mu^\ct dx^\mu=-{a^2\over x^2}{|x_1|^2d\a_1+|x_2|^2d\a_2\over 2x^2+a^2}
\label{monopotx}
\ee
and turns out to be a particular case of (A.12), 
when $P_\z$ is the $U(1)$ principal bundle over the EH manifold.
In the coordinates employed in (\ref{ehmetric}) and with a proper gauge 
choice, $A_\mu^{\ct}$ may be identified with the monopole potential 
(\ref{ultima}). In the same coordinate system,
inserting (\ref{ugauge}) in (\ref{connec}), one explicitly gets
\be
A = A_\mu dx^\m = 
i\pmatrix{f(r)\s_z & g(r)\s_- \cr 
g(r)\s_+ & -f(r)\s_z \cr}
\label{aconnec}
\ee
where $\sigma_\pm=\sigma_x\pm i\sigma_y$ and
\be
f(r)={t^2 r^2+a^4\over r^2(r^2+t^2)}\qquad g(r)={\sqrt{t^4-a^4}\over r^2+t^2}
\label{fg}
\ee
with $t^2=2s^2-a^2$ and $r^2=2x^2+a^2$.
The resulting self-dual field strength is given by
\be
F =
i\pmatrix{H(rdr\wedge\s_z+r^2\s_x\wedge\s_y)
&G({r^2\over u}dr\wedge\s_-+ur\s_+\wedge\s_z)\cr 
G({r^2\over u}dr\wedge\s_++ur\s_-\wedge\s_z)
&-H(rdr\wedge\s_z+r^2\s_x\wedge\s_y)\cr}
\label{fieldstrength}
\ee
with
\bea
H &=&{2\over r}{df\over dr}= {4 (t^2r^4+2a^4r^2+a^4t^2) \over r^4(r^2+t^2)^2}
\cr\label{funzionifs}\cr
G &=&{2u\over r^2}{dg\over dr} = {4 u \sqrt{t^4-a^4}\over r (r^2+t^2)^2}
\eea
Using (\ref{fieldstrength}), 
the second Chern class may be checked to be $1/2$, as expected.

The connection (\ref{aconnec}) was previously found in \cite{bcc}
following a completely different procedure. In the limit $a\to 0$ 
(\ref{aconnec}) becomes 
a connection over $\real^4/\zet_2$ and coincides with the BPST 
instanton (in the singular gauge) with center located at $x_0=0$ and size $t$ 
\cite{aty}.
This is no surprise since, in this limit, this construction gives the 
usual ADHM construction for $SO(3)$ bundles \cite{cws}. 

\subsection{$SU(2)$ Gauge Bundle with $ c_2({\cal E})=1$}

This case corresponds to the choice $v=(1,1),
w=(2,0), u=(2,0)$. Following the same reasoning as in the previous 
section and the results in the appendix, the matrices 
in (\ref{defaa}), (\ref{defab}) can be cast into the form
\be
A=\pmatrix{0 & a_1\cr a_2 &0\cr},\quad B=\pmatrix{0 & b_1\cr b_2 &0\cr}
\label{cinquetre}
\ee
We remark that, since $w_1=0$, the map $t^\dagger \in\Hom(V_1,W_1)$ is 
absent. Employing a notation similar to (\ref{cinqueuno}), we write the 
sections of the bundles of interest as 
\bea
\pmatrix{\vv_0^1\cr \vv_0^2}\otimes\pmatrix{0\cr \phi_1} &\in&
\bar Q\otimes \bar V_0\otimes\ct_1 \cr
\pmatrix{\vv_1^1\cr \vv_1^2}\otimes\pmatrix{\phi_{_0} \cr 0} &\in&
\bar Q\otimes \bar V_1\otimes\ct_0 \cr
\pmatrix{\w_0^1\cr \w_0^2}\otimes\pmatrix{\phi \cr 0} &\in&
\bar W_0\otimes\ct_0
\label{cinquequattro}
\eea

By eliminating rows and columns of $\cd^\dagger_{\G}$ that, when 
applied to (\ref{cinquequattro}) lead to no constraint,
the linear map (\ref{linmap}) can be restricted to the form
\be
\cd_\G=\pmatrix{y_1&-\bar x_2&a_1&-\bar b_2\cr 
y_2&\bar x_1&a_2&\bar b_1\cr
b_1&-\bar a_2&x_1&-\bar y_2\cr 
b_2&\bar a_1&x_2&\bar y_1\cr
s_1&-\bar t_2&0&0\cr 
s_2&\bar t_1&0&0\cr}
\label{cinquesei}
\ee

We start by solving the ADHM equations (\ref{adhmeq}) 
in the limit $a\to 0$.
The condition $\cd^\dagger\cd=\Delta\otimes\unita$ with $\Delta$ a 
hermitean matrix 
leads to $b_i=a_i, t_i=s_i, y_i=x_i$. The adjoint of (\ref{cinquesei}) 
can then be written as a matrix of quaternions
\be
\cd^{\dagger}_\G=\pmatrix{ x^\dagger &p^\dagger &s^\dagger \cr 
p^\dagger &x^\dagger &0\cr}
\label{cinquesette}
\ee
where $p, s$ and $x$ are quaternions whose components are the complex 
numbers $a_i,s_i$ and $x_i$ ($i=1,2$) respectively. To compute the kernel  
of the matrix (\ref{cinquesette}) we rewrite it 
as $\cd_\G^\dagger=(D,q)$ \cite{cws}, where 
\be
q=\pmatrix{s^\dagger\cr 0\cr} \qquad D=\pmatrix{x^\dagger &p^\dagger\cr 
p^\dagger &x^\dagger\cr}
\label{qmquater}
\ee
Notice that $q$ and $D$ are respectively the restrictions of the maps 
$\Psi^\dagger$ and ${\cal A}^{\dagger}$, defined in (\ref{defab}) and 
(\ref{defaa}). 
The reorganization of the ADHM data 
in quaternionic notation is always possible for $a\to 0$.
In fact, as we have already noticed, in this limit the construction 
we are 
describing becomes the usual ADHM construction on the orbifold 
$\real^4/\zet_2$.
In the same vein, we rewrite the frame $U$ in (\ref{connec}) 
as $U=\pmatrix{\vv\cr \u\cr}$, so that the
matrix $q$ ($D$) will be acting on the quaternion $\u$ ($\vv$). 

The condition 
$\cd_\G^\dagger U=0$ yields 
\be
\vv=-D^{-1}q\u
\label{cinqueotto}
\ee
and $U=\pmatrix{-D^{-1}q\u\cr \u\cr}$.
The unitarity constraint  $U^\dagger U =\unita$ then gives
\be
\u =
{1\over \sqrt{1+q^\dagger(DD^\dagger)^{-1}q}}\tilde\u
\label{cinquenove}
\ee
with $\tilde\u^\dagger\tilde\u =\unita$. By an $SU(2)$ gauge transformation 
one can always set $\tilde\u=\unita$. Exploiting this freedom, 
(\ref{connec}) can be written in the form
\be
A_\mu=\mezzo \u^2[(D^{-1}q)^\dagger\de_\mu(D^{-1}q)-(\de_\mu
(D^{-1}q)^\dagger)(D^{-1}q)]
\label{cinquedieci}
\ee
The only delicate point in these formulae is the computation of 
the matrix $D^{-1}$ with the property $D^{-1}D=1$. As $D$ is a 
matrix
of quaternions $D$ does not commute with $D^{-1}$. The left 
and the right
inverse of $D$ are not necessarily equal.
To compute the left inverse, 
$D^{-1}$, we observe that it is always possible to find matrices
$E_k$ such that $E_k\ldots E_2E_1D=\unita$ \cite{fin}.
$E_k$ is a matrix representing one of the three possible elementary 
operations on rows
\begin{itemize}
\item [{i)}]
P: permutation of two rows
\item [{ii)}]
A: addition of two rows
\item [{iii)}]
M: multiplication of a row by a number
\end{itemize}
A matrix can be diagonalized by a finite number of P, A, M 
operations.
Following the appropriate steps on the matrix $D$, one finds
\be
D^{-1}={1\over\Phi}\pmatrix{(x^2+p^2) x^\dagger-
2(x\cdot p)p^\dagger
& (x^2+p^2)p^\dagger-2(x\cdot p)x^\dagger\cr
(x^2+p^2)p^\dagger-2(x\cdot p)x^\dagger 
&(x^2+p^2)x^\dagger-2(x\cdot p)p^\dagger}
\label{cinqueundici}
\ee
where $\Phi=(x-p)^2(x+p)^2$ and 
\be
U={1\over\sqrt{1+{p^2\over(x+p)^2}+{p^2\over(x-p)^2}}}
\pmatrix{1\cr {p^\dagger\over\sqrt{(x-p)^2}}\cr 
{p^\dagger\over\sqrt{(x+p)^2}}}
\label{cinquedodici}
\ee
In (\ref{cinqueundici}) $x\cdot p=\mezzo(x^\dagger p + p^\dagger x)$ 
is the scalar product of the two vectors $x^\m$ and $p^\m$, 
\ie~$\mezzo(\bar x_1 a_1+\bar x_2 a_2+\bar a_1 x_1+
\bar a_2 x_2 = x^\m p_\m$. As expected,
the resulting gauge connection $A_\mu^a=-i
\eta^a_{\mu\nu}\de^\nu \ln \u^2$ is invariant 
under the $\zet_2$ transformation $x\to -x$.

We are now ready to solve the KN-ADHM equations 
for $a\neq 0$. Going back to (\ref{cinquesei}), we write
\be
\cd^\dagger_\G=
\pmatrix{\bar y_1 & \bar y_2 & \bar b_1 & \bar b_2 
& \bar s_1 &\bar s_2\cr
-x_2 & x_1 & -a_2 & a_1 & -t_2 & t_1\cr
\bar a_1 & \bar a_2 & \bar x_1 & \bar x_2& 0 & 0\cr
-b_2 & b_1 & -y_2 & y_1 & 0 & 0}
\label{cinquedodicibis}
\ee
The condition $\cd^\dagger\cd=\Delta\otimes\unita$ with $\Delta$ a 
hermitean matrix implies $t_i=s_i$, $y_i=\lambda x_i$ and 
$b_i=\mu a_i$, where $\lambda^2 = 1 + a^2/x^2$ and 
$\m^2=1-a^2/p^2$ can be chosen to be real. As before $p$ is the 
quaternion made out of the two complex numbers $a_1, a_2$.
It is again convenient to put $\cd^\dagger_\G=(D,q)$, 
where $D$ is a $4\times4$ 
complex matrix and $q^\dagger=(s,0)$ with $s$ the 
quaternion made out of $s_1, s_2$.
The frame of sections $U$ can be put into the form
\be
U=\pmatrix{\vv_1 & \vv_2 \cr\u_1 & \u_2}
\label{cinquetredici}
\ee
where $\vv_1, \vv_2$ and $\u_1, \u_2$ are four-dimensional 
and two-dimensional complex vectors, respectively. 
The constraint (\ref{trequattro}) becomes
\be
D\vv_i=-\pmatrix{s^\dagger\u_i\cr 0}
\label{cinquequattordici}
\ee
and yields for $\vv_i$
\be
\vv_i=D^\dagger(D D^\dagger)^{-1}\pmatrix{s^\dagger\u_i\cr 0}
\label{cinquequindici}
\ee
Since $DD^\dagger$ turns out to be made of real quaternions only, 
the inversion of this operator is easily performed and gives
\be 
(D D^\dagger)^{-1}=
{1\over\Delta}\pmatrix{(x^2+p^2)\otimes\unita&-c\otimes\unita\cr 
\bar c\otimes\unita & (x^2 +p^2)\otimes\unita\cr} 
\label{cinquesed}
\ee
where 
\bea
c&=&\lambda(a_1\bar x_1+a_2\bar x_2)+\mu (x_1\bar a_1+x_2\bar a_2)\cr
\Delta&=&(x^2+p^2)^2-|c|^2
\label{candd}
\eea
Imposing the unitarity condition
$U^\dagger U=\unita$ (\ie~$\u^\dagger\u + \vv^\dagger\vv=\unita$) yields
\be
\u=\sqrt{\Delta\over \Delta + s^2(x^2+p^2)}\tilde\u
\label{cinquedicias}
\ee
with $\tilde\u^\dagger\tilde\u=\unita$.
Exploiting as before the $SU(2)$ gauge symmetry one can set 
$\tilde\u=\unita$,
finally getting
\be
\vv_1 = {1\over\Lambda}
\pmatrix{(x^2+p^2)(\lambda x_1\bar s_1+s_2\bar x_2)-
\bar c(a_1\bar s_1+\mu s_2\bar a_2)\cr
(x^2+p^2)(\lambda x_2\bar s_1-s_2\bar x_1)-
\bar c(a_2\bar s_1-\mu s_2\bar a_1)\cr
(x^2+p^2)(\mu a_1\bar s_1+s_2\bar a_2)-
\bar c(x_1\bar s_1+\lambda s_2\bar x_2)\cr
(x^2+p^2)(\mu a_2\bar s_1-s_2\bar a_1)-
\bar c(x_2\bar s_1-\lambda s_2\bar x_1)\cr}
\label{cinquediciota}
\ee
\be
\vv_2 = {1\over\Lambda}
\pmatrix{(x^2+p^2)(\lambda x_1\bar s_2-s_1\bar x_2)-
\bar c(a_1\bar s_2-\mu s_1\bar a_2)\cr
(x^2+p^2)(\lambda x_2\bar s_2+s_1\bar x_1)-
\bar c(a_2\bar s_2+\mu s_1\bar a_1)\cr
(x^2+p^2)(\mu a_1\bar s_2-s_1\bar a_2)-
\bar c(x_1\bar s_2-\lambda s_1\bar x_2)\cr
(x^2+p^2)(\mu a_2\bar s_2+s_1\bar a_1)-
\bar c(x_2\bar s_2+\lambda s_1\bar x_1)\cr}
\label{cinquediciotb}
\ee
where $\Lambda=\sqrt{\Delta [\Delta + s^2(x^2+p^2)]}$.

Together with (\ref{cinquedicias}), (\ref{cinquediciota}) and 
(\ref{cinquediciotb}) allow to compute $U$ and consequently 
the gauge connection. As expected, the solution depends on 
eight real parameters. These are the four complex numbers 
$s_1, s_2, a_1$ and $a_2$.

\subsection{$SU(2)$ Gauge Bundle with $ c_2({\cal E})={3\over 2}$}

This is the case corresponding to the deformation of the
solution studied in \cite{bfmr} 
which is obtained from the standard embedding of the 
spin into the gauge connection.
This solution is identified by the choice $w=(0,2), u=(2,0),
v=(1,2)$. (\ref{defaa}) is a $2\times 2$ matrix of 
the form (\ref{cinquetre}), 
but the entries are themselves matrices describing the maps 
$\Hom (V_i,V_j)$. Since $v_1=1$ and $v_2=2$, these are 
$2\times 1$ and 
$1\times 2$  matrices. Therefore the complex matrices $A, B$ are of 
the form
\be
A=\pmatrix{0 & a_1 & a_2\cr a_3 & 0 & 0\cr a_4 & 0 & 0}, \qquad
B=\pmatrix{0 & b_1 & b_2\cr b_3 & 0 & 0\cr b_4 & 0 & 0}
\label{cinquesedici}
\ee

The maps $s, t^\dagger$ in (\ref{defab}) 
belong to $(\oplus_i\Hom(W_i,V_i))_{\G}$ and can be
represented by the $2\times3$ matrices
\be
s=\pmatrix{0 & s_1 & s_2\cr
0 & s_3 & s_4},\qquad
t^\dagger=-\pmatrix{0 & \bar t_1 & \bar t_2\cr
0 & \bar t_3 & \bar t_4}
\label{cinquediciasette}
\ee 
The zero entries correspond to maps involving the null space $W_0$, 
which has dimension $w_0=0$.

The section $u_{_U}$ can be written in this case
\be
u_{_U}=\pmatrix{\vv_0^1\otimes\pmatrix{0\cr\phi_1^1}\cr
\pmatrix{ \vv^1_1\cr \vv^1_2}\otimes\pmatrix{\phi^1_0\cr 0}\cr 
\vv_0^2\otimes\pmatrix{0\cr\phi_2^1}\cr
\pmatrix{ \vv^2_1\cr \vv^2_2}\otimes\pmatrix{\phi_2\cr 0}\cr
\pmatrix{\w_1\cr \w_2}\otimes\pmatrix{0\cr \phi}}
\label{cinquediciotto}
\ee
The action of the term $\unita\otimes\xi^\dagger$, 
in (\ref{defd}), on $u_{_U}$ 
is made more transparent if we assemble the first four rows of 
(\ref{cinquediciotto}) into one doublet
\be
\pmatrix{\Sigma\cr\Lambda\cr}\equiv\pmatrix{\vv_0^1\otimes
\pmatrix{0\cr\phi_1^1}\oplus
\pmatrix{\vv^1_1\cr \vv^1_2}\otimes\pmatrix{\phi^1_0\cr 0}\cr
\vv_0^2\otimes\pmatrix{0\cr\phi_2^1}\oplus
\pmatrix{ \vv^2_1\cr \vv^2_2}\otimes\pmatrix{\phi_2\cr 0}\cr}
\label{cinquediciannove}
\ee
Consequently
\be
(\unita\otimes\xi^\dagger) \pmatrix{\Sigma\cr\Lambda\cr}=
\pmatrix{\alpha^\dagger\Sigma+\beta^\dagger\Lambda\cr 
-\beta\Sigma+\alpha\Lambda\cr}
\label{nsbdk}
\ee
Neglecting rows and columns of $\cd_\G^\dagger$ that when acting on 
(\ref{cinquediciotto}) lead to no constraint, the expression of 
$\cd_\G^\dagger$ can be reduced to 
\be
\cd_\G^\dagger=\pmatrix{
\bar y_1&\bar a_3&\bar a_4&\bar y_2&\bar b_3&\bar b_4&0&0\cr
\bar a_1&\bar x_1&0&\bar b_1&\bar x_2&0&\bar s_1&\bar s_3\cr
\bar a_2&0&\bar x_1&\bar b_2&0&\bar x_2&\bar s_2&\bar s_4\cr
-x_2&b_1&-b_2&x_1&a_1&a_2&0&0\cr
-b_3&y_2&0&a_3&y_1&0&t_1&t_3\cr
-b_4&0&y_2&a_4&0&y_1&t_2&t_4\cr}
\label{cinqueventi}
\ee
In the limit $a\to 0$ the ADHM equations (\ref{adhmeq}) 
are satisfied by
\bea
&\,& y_1=x_1 \quad y_2=x_2\cr
&\,& a_3=a_1,\quad a_4=a_2,\quad b_3=b_1,\quad b_4=b_2\cr
&\,& s_1={a_2\over c},\quad s_2=a_1 c,\quad
s_3={b_2\over c},\quad  s_4=b_1c 
\label{cinqueventuno}
\eea
with $c$ an arbitrary real number. By a reshuffling of rows and columns 
the map (\ref{cinqueventi}) can 
be rewritten as a matrix of quaternions
\be
\cd_\G^\dagger=\pmatrix{x&q_1&q_2&0\cr q_1&x&0&{q_2/c}\cr
q_2&0&x&q_1 c}=(D,q)
\label{cinqueventidue}
\ee
where $x$, $q_1$ and $q_2$ are quaternions whose components are the complex
numbers $x_i$, $a_i$ and $b_i$ ($i=1,2$), respectively. $D$ is the 
$3\times3$ left block of $\cd_\G^\dagger$. Exploiting the quaternionic 
notation, we write the frame of sections $U$ as
\be
U=\pmatrix{\vv \cr \u}
\label{cinquetredicib}
\ee
where $\vv$ and $\u$ are three-dimensional 
and one-dimensional quaternions respectively. 
The constraint (\ref{trequattro}) becomes
\be
D\vv=-\pmatrix{s^\dagger\u\cr 0}
\label{cinquequattordicib}
\ee
and yields for $\vv$
\be
\vv=-D^{-1} q^\dagger\u
\label{cinquequindicibis}
\ee

The solution can be expressed in terms of the left inverse of 
$D$, $D^{-1}$, whose matrix elements are computed to be 
\bea
D^{-1}_{11}&=& {1\over\Delta}
x^2[(x^2+q_1^2+q_2^2)\bar x-2(x\cdot q_2)\bar q_2-2 
(x\cdot q_1)\bar q_1]\cr
D^{-1}_{12}&=& {1\over\Delta}
[-2(x\cdot q_1)(x^2+q_2^2)\bar x+x^2(x^2+q_1^2+q_2^2)\bar q_1
+2(x\cdot q_2)\bar q_2q_1\bar x]\cr
D^{-1}_{13}&=&{1\over\Delta}
[-2(x\cdot q_2)(x^2+q_1^2)\bar x+x^2(x^2+q_1^2+q_2^2)\bar q_2
+2(x\cdot q_1)\bar q_1q_2\bar x]\cr
D^{-1}_{21}&=&{1\over\Delta}
[-2(x\cdot q_1)(x^2+q_2^2)\bar x+x^2(x^2+q_1^2+q_2^2)\bar q_1
+2(x\cdot q_2)\bar x q_1\bar q_2]\cr
D^{-1}_{22}&=&{1\over\Delta}
[(x^2+q_1^2+q_2^2)(x^2+q^2_2-4(x\cdot q_2)^2)\bar x
+(4(x\cdot q_2)(q_1\cdot q_2)\cr 
&&-2x\cdot q_1(x^2+q_2^2))\bar q_1-2(x\cdot q_2)q_1^2\bar q_2]\cr
D^{-1}_{23}&=&{1\over\Delta}
[4(x\cdot q_2)(x\cdot q_1)\bar x-2(x\cdot q_2)(x^2+q_1^2)\bar q_1
+2(x\cdot q_1)q^2_1\bar q_2\cr
&&-(x^2+q_1^2+q_2^2)\bar x q_1\bar q_2]\cr
D^{-1}_{31}&=&{1\over\Delta}
[-2(x\cdot q_2)(x^2+q_1^2)\bar x+x^2(x^2+q_1^2+q_2^2)\bar q_2
+2(x\cdot q_1)\bar x q_2\bar q_1]\cr
D^{-1}_{32}&=&{1\over\Delta}
[4(x\cdot q_2)(x\cdot q_1)\bar x-2(x\cdot q_1)(x^2+q_2^2)\bar q_2
+2(x\cdot q_2)q^2_2\bar q_1\cr
&&-(x^2+q_1^2+q_2^2)\bar x q_2\bar q_1]\cr
D^{-1}_{33}&=&{1\over\Delta}
[(x^2+q_1^2+q_2^2)(x^2+q^2_1-4(x\cdot q_1)^2)\bar x
+(4(x\cdot q_1)(q_1\cdot q_2)\cr &&-2(x\cdot q_2)(x^2+q_1^2))\bar q_2
-2(x\cdot q_1)q_2^2\bar q_1]
\label{cinqueventitre}
\eea
with 
\be
\Delta=(x^2+q_1^2+q_2^2)-4(x^2+q_1^2)(x\cdot q_2)^2
-4(x^2+q_2^2)(x\cdot q_1)^2+8x\cdot q_1x\cdot q_2q_1\cdot q_2
\label{cinqueventiquattro}
\ee
The last step of this long calculation is to fix the form
of the quaternion $\u$ by the orthogonalization condition 
(\ref{trequattrobis}) and the symmetries of the construction.
We do not report here the complicated, uninspiring expression 
of $\u$. 

As expected the solution depends on twelve parameters. 
The eight real parameters of the two quaternions $q_1, q_2$, the real 
constant $c$ and three angles associated with global $SU(2)$ rotations.
The latter are left implicit in the construction.
The $a\neq 0$ case turns out to require quite a formidable number of
algebraic manipulations. As our main
interest lies in the computation of instanton 
dominated correlators for (locally) SUSY gauge theories, the 
$a=0$ solution 
appears to be sufficient for the  $N=1,2$ cases, as remarked 
in \cite{bbfmr}.
In this respect, it is interesting to note that the two-center 
solution derived from (\ref{bccth}) and (\ref{covlaplace}) has 
the right number of parameters to match the dimension of the 
moduli space
for the case $\kappa=3/2$. 

\section{Moduli Spaces of Gauge Connections of Yang-Mills 
instantons on ALE manifolds}
\setcounter{equation}{0}
The KN construction establishes an isomorphism between the 
equivalence class of solutions of the ADHM equations on ALE spaces, $X$,
and the moduli space of gauge connections on instanton bundles, ${\cal E}$. 
It can be proved that this isomorphism is an \hk isometry 
and, as we already argued, the moduli space,
${\cal M}_{\cal E}$, turns out to be a \hk manifold \cite{kn}
(for a pedagogical introduction to the structure of moduli spaces see
\cite{hita}).
The metric $G$ on ${\cal M}_{\cal E}$ is given by
\be
G_{IJ} = \int_X d^4x \sqrt{{\rm det}(g)} g^{\mn}
\delta_IA_\m^i \delta_JA_\n^j \delta_{ij}
\label{modulimetric}
\ee 
where $g$ is the metric on $X$ and $\delta A$ are the zero-mode
fluctuations of the gauge fields around the instanton configuration.
The three \hk forms on $X$ can also be mapped into the moduli space, 
${\cal M}_{\cal E}$, by the formulae
\be
\W^a_{IJ} = \int_X d^4x \omega^a_{\m\n}
\delta_IA_i^\m \delta_JA^\n_j \delta^{ij}
\label{moduliform}
\ee 
The three resulting two-forms, $\W^a$, are closed and thus define a 
hyper-K\"ahler structure on the moduli space \cite{hita}.

There is one case in which ${\cal M}_{\cal E}$ is completely known. 
To show this we have to use a very powerful theorem proven in 
\cite{kn, nak}, 
that states that on any ALE manifold, $X$, there always exists 
a bundle, ${\cal E}$, 
with $c_1({\cal E})=0, c_2({\cal E})=(|\G|-1)/|\G|$
such that the moduli space, ${\cal M}_{\cal E}$ is 
four-dimensional and coincides with the base manifold 
$X$ itself. Furthermore there exists a point
$m_0\in {\cal M}_{\cal E}$ at which 
${\cal E}(m_0)={\cal L}\oplus\bar{\cal L}$ with 
${\cal L}$ a line bundle, implying that ${\cal E}$ is reducible at $m_0$. 
While on compact manifolds $m_0$ must be a singular point of the moduli 
space \cite{fu, dk}, this is not necessarily true on non-compact manifolds 
\cite{nak}.
For EH manifold with metric (\ref{ehmetric}) $r=a$ is such a point.
Going back to our previous discussion and using this theorem, we see 
that the moduli space of connections on the minimal $SU(2)$ instanton
(the one with $\kappa={1/2}$) has dimension four (from (\ref{quattrosei})) 
and precisely coincides with the EH manifold. Here we have 
a nice example of symmetry enhancement at a singular point of a 
moduli space. As depicted in Fig.2,
the EH manifold is the smooth resolution of the singularity of the
orbifold $\complex^2/\zet_2$: the parameter $a$ appearing in 
(\ref{ehmetric})  
measures the distance from the singularity. In the limit 
$a\mapsto 0$ the $SU(2)\times U(1)$ symmetry of the moduli space 
gets enhanced to $SU(2)\times SU(2)$. 
\vskip .6cm
\centerline{\vbox{\epsfysize=40mm \epsfbox{moduli.eps}}}
\smallskip
\vskip .4cm
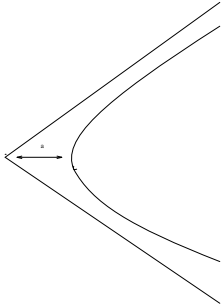
\centerline{\bf Figure 2}
\vskip .6cm

In order to explicitly compute (\ref{modulimetric}), 
one needs the four zero-modes fluctuations 
of the gauge fields around the minimal instanton. 
They are related to the global symmetries broken by the 
instanton background, \ie~dilatations and $SU(2)$ rotations. Notice that
translation \zms do not exist because
the $\zet_2$ identification $x \approx -x$ forbids global 
translations in the 
EH manifold \cite{eh}. 
We must look for zero modes that satisfy the background gauge 
condition
and that, up to a local term of the form $D_\m \Lambda$ \cite{yabe}, have 
the form of derivatives of the gauge connection with respect 
to the four free
parameters (collective coordinates) \cite{acto} appearing in 
(\ref{aconnec}).

For the zero-mode associated to dilatations the situation is very simple, 
because the derivative of (\ref{aconnec}) with respect to $t$, \ie
\be
\delta_0 A = {\de A \over \de t} = {i t \over \sqrt{t^4-a^4}}
\pmatrix{f^3(r)u\s_z & f^1(r)r\s_- \cr 
f^1(r)r\s_+ & -f^3(r)u\s_z \cr}
\label{dilatation}
\ee
with 
\be
f^3(r)={2(t^2r^2+a^4)\over r(r^2+t^2)^2} \quad
f^1(r)={2\sqrt{r^4-a^4}\sqrt{t^4-a^4}\over r(r^2+t^2)^2} 
\label{trasvff}
\ee
automatically satisfies the background gauge condition, 
$D^\m \delta_0 A_\m=0$. 
The zero-modes associated to global $SU(2)$ rotations 
\be
{\de A^{\W(\theta_{_0})} \over \de \theta^j_{_0}} 
\Big |_{\theta_{_0}=0} = i [T_j,A]
\label{rotation}
\ee
are not transverse. In order to make them transverse one has to 
add a local term of the form $D_\m \Lambda_j$ \cite{yabe}. Putting
$\Lambda_j = (\theta_j - \theta_j^{(o)})$, the $\theta_j$ turn out to 
be the bounded harmonic scalars given by \cite{abfmr, bbfmr}
\be
\theta_3={t^2r^2+a^4\over t^2(r^2+t^2)} \theta_3^{(o)}, \quad
\theta_{1,2}={\sqrt{r^4-a^4}\over r^2+t^2} \theta_{1,2}^{(o)} 
\label{scalzero}
\ee
The resulting three transverse zero-modes have the final expression
\be
\delta_j A^r_\m = (D_\m \theta_j)^r
\label{rotrans}
\ee
The four transverse zero-modes $\delta_I A$ form an orthogonal basis with 
norms 
\bea
&\,& ||\delta_0 A||^2={8\pi^2 \over g^2}{t^4 \over t^4-a^4} \cr
&\,& ||\delta_1 A||^2=||\delta_2 A||^2={8\pi^2 \over g^2} t^2\cr
&\,& ||\delta_3 A||^2={8\pi^2 \over g^2}{t^4-a^4 \over t^2}
\label{norme}
\eea
where we have introduced back the gauge coupling constant, $g$.

The framed moduli space of the \min is 
four-dimensional and locally looks like $\real^+ \times S^3/\zet_2$,
with $t$ ($t \ge a$) playing the role of the radial variable.
Near the identity element of $SU(2)$ the metric is given by
\be
G_{IJ} = \int_X d^4x \sqrt{det(g_{_{EH}})} g^{\mn}_{_{EH}}
\delta_IA_\m^r \delta_JA_\n^s \delta_{rs}
\label{minmodulimet}
\ee 
The metric $G_{IJ}$ may be transported to any element of the group 
$S^3/\zet_2$
by left translations and one may check that it is identical (up to an
overall rescaling) to the metric in (\ref{ehmetric}), after the
substitutions 
\be
t\rightarrow r, \qquad
exp(i\theta^kT_k) \rightarrow exp(i\psi T_3) exp(i\theta T_2) exp(i\phi T_3)
\label{finale}
\ee

\section{Conclusions}

In this paper we have tackled the problem of the explicit construction
of self-dual gauge connections on ALE manifolds and we have 
carried 
out computations for the simplest of all these manifolds the 
EH gravitational instanton. The results obtained here should allow to 
generalize 
the computations of instanton dominated correlation functions 
performed in
\cite{bbfmr}. There we computed the gaugino condensate in a globally 
$N=1$ supersymmetric YM theory on the
background of the EH manifold by expanding the functional
integral around the minimal $SU(2)$ instanton and we found a constant 
finite
value as expected on supersymmetry grounds.
The explicit calculation of instanton effects in a 
supersymmetric YM theory
coupled to supergravity seems to be viable. The relevant Green 
functions
contain a gravitational sector (represented by the field strength 
gravitino bilinear)
and a gauge sector (represented by the gaugino bilinears) which 
are completely 
decoupled
as argued in \cite{bbfmr}. More explicitly, the result of the 
functional 
integration in the gauge sector is independent from 
the moduli of the gravitational background, 
$\zeta_\ccomplex$ and $\zeta_\rreal$. This observation suggests that it may 
suffice to solve the ADHM equations on ALE spaces in the orbifold limit, 
$\zeta_\rreal=0, \zeta_\ccomplex=0$.
This, as we have seen, is a great simplification and it 
is the reason why we have presented
separately many of the results for the EH background in the limit 
$a\to 0$.
Unfortunately, even in this limit, the final form of the $SU(2)$ 
gauge connection which corresponds to the standard
embedding solution, \ie~the one with $c_2({\cal E})=3/2$ and 
$\dim({\cal M}_{\cal E})=12$, looks quite involved, making the 
computation
of the relevant zero-modes very complicated. A possible way out 
could be to perform these calculations by resorting to the two-center 
ansatz (\ref{bccth}), (\ref{covlaplace}).

It is interesting to note that 
the multi-center ansatz for self-dual $SU(2)$ gauge connections 
seems to 
work for any four-dimensional hyper-K\"ahler manifold, since it 
essentially 
relies
on the existence of three covariantly constant complex structures. 
It may then
prove to be useful in the study of certain properties of the moduli 
spaces of YM instantons on the $K3$ surface, which seem to play a 
fundamental role
in the issue of string-string duality in six dimensions
and in the intriguing relation between $K3$ and ALE instantons.

As a side remark we have presented the computation of the 
partition sum for an 
abelian theory with a $\theta$-term, outlining the lack of 
S-invariance on ALE 
spaces, at least in
the form which works for other compact hyper-K\"ahler manifolds 
\cite{ver,
wittab}. Still, the modular properties of the level-one characters of the 
affine A-D-E Lie algebras, appearing in the abelian partition 
function, as well as those of higher level, appearing in non-abelian  cases 
\cite{vw}, call for some deeper explanation.

\appendix
\section{ \large\bf Appendix}
\setcounter{equation}{0}
\setcounter{section}{1}
The hyper-K\"ahler quotient construction of the EH metric starts from 
$\Xi=\complex^4$ endowed with the flat metric 
\be
ds^2_{\Xi}=\sum_{i=1}^2(|dx_i|^2+|dy_i|^2) 
\label{duetredicia}
\ee
The metric (\ref{duetredicia}) and the \hk two-forms
\be
\omega_\ccomplex = \sum_{i=1}^2 dx_i \wedge dy_i \qquad
\omega_\rreal =  \sum_{i=1}^2(dx_i\wedge d\bar x_i + dy_i
\wedge d\bar y_i) 
\label{ehforme}
\ee
are invariant under
\be
x_i \mapsto e^{i\omega}x_i\qquad
y_i \mapsto e^{-i\omega}y_i
\ee
The moment maps relative to this $U(1)$ triholomorphic isometry
are
\be
\m_{\ccomplex} = x_1 y_1 - x_2 y_2
\qquad \m_{\rreal}=\sum_{i=1}^2(|x_i|^2-|y_i|^2) 
\label{ehmoment}
\ee
Correspondingly the five-dimensional submanifold 
$P_{\z} = \{\xi\in \Xi  \, \vert \; \m_{\rreal}(\xi) = \z_{\rreal}, 
\m_{\ccomplex}(\xi)=0\}$ is determined by
\bea
0=\z_\ccomplex &=&[\a,\b]\cr
a^2=\z_\rreal &=&[\a,\a^\dagger]+[\b,\b^\dagger]
\label{duetrediciaa}
\eea
where $\a, \b$ are defined in (\ref{questaserve}).
The EH instanton is the simplest of the ALE manifolds and corresponds to
the choice $\G=\zet_2$. 
The condition of $\G$-invariance on the doublet of matrices 
$(\a, \b)$ reads 
\be
\pmatrix{\rho_{_R}(\g) \a \rho_{_R}(\g^{-1}) \cr
\rho_{_R}(\g) \b \rho_{_R}(\g^{-1})\cr} = \rho_{_Q}(\g) 
\pmatrix{\a\cr \b\cr}
\label{duequat}
\ee
The matrices which satisfy the constraint (\ref{duequat}) are $2\times 2$ 
matrices with only (complex) off-diagonal entries.
Equations (\ref{duetrediciaa}) are solved by $y_1=
\lambda x_2, y_2=\lambda x_1$, where $\lambda$ is a complex number 
with $|\lambda|^2=1+{a^2/x^2}$ and $x^2=\sum_i |x_i|^2$. The metric on 
the submanifold $P_{\z}$ then becomes
\be
ds^2_{P_\z}=\sum_{i=1}^2(|dx_i|^2+|\lambda|^2|dx_i+x_i
{d\lambda\over\lambda}|^2)
\label{duequindici}
\ee
from which, putting 
$\lambda=|\lambda| e^{i\b}$ and $x_i=r_ie^{i\a_i}$, one obtains
\be
ds^2_{P_\z}=\sum_{i=1}^2[(dr_i^2+(|\lambda|dr_i+r_id|\lambda|)^2)+
r_i^2(d\a_i^2
+|\lambda|^2(d\a_i+d\b)^2)]
\label{duesedici}
\ee
It is convenient to divide the line element into two parts 
$ds^2_{P_\z}\equiv 
ds^2_r+ds^2_\Omega$, with $ds^2_r$ containing the 
differentials of the variables $r_i, |\lambda|$ and $ds^2_\Omega$ the 
differentials of $\a_i, \b$. In fact $ds^2_{\Omega}$ has still a 
residual $G=U(1)$ invariance associated to a combination of the 
variables $\a_i, \b$. 
The Killing vector related to the corresponding cyclic variable is
\be
k=k^\mu\de_\mu=\sum_i\bigl({\de\over\de\a_i}-{\de\over\de\b}\bigr)
\label{duediciasette}
\ee

The next step is to construct $X_\z={P_\z/G}$. As it is nicely
explained in \cite{hklr}, from a physicist's point of view, 
this construction
can be regarded as a gauging procedure of a non-linear $\sigma$-model on 
$\real^d$ with Lagrangian density
\be
L=\mezzo g^{P}_{\mu\nu}(\phi)\de^a\phi^\mu\de_a\phi^\nu
\label{duediciotto}
\ee
where $g^{P}_{\mu\nu}(\phi)$ is the metric on the target space $P$ 
with coordinates 
$\phi^\mu$ and $\de_a={\de/\de x^a}$, with $x^a$, $a=1,\ldots,d$, 
coordinates 
on $\real^d$. Let $P$ admit a group of isometry $G=U(1)$ 
whose action on the 
fields $\phi^\m$ is
\be
\delta\phi^\mu=\epsilon k^\mu(\phi)
\label{duediciannove}
\ee
with $\epsilon$ constant. The global symmetry (\ref{duediciannove}) can be
promoted to a local one by introducing the connection
\be
A_a=-{\de_a\phi^\mu g^P_{\mu\nu}k^\nu\over g^P_{\r\s}k^\r k^\s}
\label{dueventi}
\ee
and substituting standard derivatives with covariant ones. 
The resulting gauged $\sigma$-model has target space $X=P/G$, 
the space of $G$ orbits on $P$ with $\dim(X)=\dim(P)-\dim(G)$ \cite{hklr}.
 
In the case under consideration $\m$ runs from 1 to $\dim(P)=5$  
and the coordinates on $P$ may be taken to be 
$\phi^\m=(r_1,r_2,\a_1,\a_2,\b)$. The global isometry of $ds^2_{P_\z}$ 
is generated by the Killing vectoc (\ref{duediciasette}), with 
$k^\m=(0,0,1,1,-2)$. 
From the expression of the gauged version of the lagrangian
(\ref{duediciotto}) one can read off the metric on $X=P/G$, namely
\be
g_{\mu\nu}^{X}=g^P_{\mu\nu}-{g^P_{\mu\rho}g^P_{\nu\sigma}k^\rho k^\sigma
\over g^P_{\lambda\tau}k^\lambda k^\tau}
\label{dueventuno}
\ee
where $g^P_{\mu\nu}$ is the metric tensor in (\ref{duesedici}). 
By construction the metric (\ref{dueventuno}) is transverse to the Killing 
vector and the local $U(1)$ isometry allows one to gauge $\b$ to zero.

Computing the connection (\ref{dueventi}) and comparing it with 
(\ref{ultima}) suggest the further change of variables
\be
\a_1 ={\psi+\varphi\over 2}, \qquad
\a_2 ={\psi-\varphi\over 2}
\label{dueventiquattro}
\ee

As for the second part of the metric, $ds^2_r$, after the 
change of variables
\bea
r_1 &=&\sqrt{r^2-a^2\over 2}\cos{\theta\over 2}\cr
r_2 &=&\sqrt{r^2-a^2\over 2}\sin{\theta\over 2}
\label{dueventidue}
\eea
it can be written in the more familiar form
\be
ds^2_r={(dr)^2\over\unoar}+{r^2\over 4}(d\theta)^2
\label{dueventitre}
\ee
Collecting all terms, we get the EH metric (\ref{ehmetric}).

To summarize we give the complete changes of variables from 
the original metric (\ref{duetredicia}) to the variables employed in 
(\ref{ehmetric})
\bea
x_1 &=&\sqrt{r^2-a^2\over 2}\cos{\theta\over 2}
e^{i{\psi+\varphi\over 2}}\quad
y_1 =\sqrt{r^2+a^2\over 2}\sin{\theta\over 2}
e^{i{\psi-\varphi\over 2}}\cr
x_2 &=&\sqrt{r^2-a^2\over 2}\sin{\theta\over 2}
e^{i{\psi-\varphi\over 2}}\quad
y_2 =\sqrt{r^2+a^2\over 2}\cos{\theta\over 2}
e^{i{\psi+\varphi\over 2}}
\label{dueventisei}
\eea


\newpage
\begin{thebibliography}{99}

\bibitem{egh}{T.~Eguchi, P.B.~Gilkey and A.J.~Hanson, Phys. Rep. {\bf 66C} 
(1980) 213.}

\bibitem{bfmr}{M.~Bianchi, F.~Fucito, M.~Martellini, G.C.~Rossi, 
Nucl. Phys. {\bf B440} (1995) 129.}

\bibitem{dkl}{M.G.~Duff, R.R.~Khuri, J.X.~Lu, {\it String Solitons},
Phys. Rep. {\bf 259C} (1995) 231.}

\bibitem{abfgz}{D.~Anselmi, M.~Billo', P.~Fre', L.~Girardello, 
A.~Zaffaroni, 
Int. Jou. Mod. Phys. {\bf A9} (1994) 3007.}
 
\bibitem{kkl}{E.~Kiritsis, K.~Kounnas and D.~L\"ust, Int. Jou. Mod. Phys.
{\bf A9} (1994) 1361.}

\bibitem{bbfmr}{M.~Bianchi, F.~Fucito, M.~Martellini, G.C.~Rossi, 
Phys. Lett.
{\bf B359} (1995) 56.}

\bibitem{bcc}{H.~Boutaleb-Joutei, A.~Chakrabarti, A.~Comtet, Phys. Rev. 
{\bf D21}(1980) 979, ibid. 2285, ibid. 2280.}

\bibitem{akmrv}{V.~Novikov, M.~Shifman, A.~Vainshtein and 
V.~Zakharov, Nucl. Phys. {\bf B229} (1983) 381, 407; {\bf B260} (1985) 157;
G.C.~Rossi and G.~Veneziano, Phys. Lett. {\bf B138} (1984) 195;
I.~Affleck, M.~Dine and N.~Seiberg, Nucl. Phys. {\bf B256} (1985) 557;
D.~Amati, K.~Konishi, Y.~Meurice, G.C.~Rossi and 
G.~Veneziano, Phys. Rep. {\bf 162C} (1988) 169.} 

\bibitem{kn}{P.B.~Kronheimer and H.~Nakajima, Math. Ann. {\bf 288} 
(1990) 263.}

\bibitem{nak}{H.~Nakajima, Invent. Math. {\bf 102} (1990) 267.}

\bibitem{gn}{T.~Gocho and H.~Nakajima, J. Math. Soc. Jap. {\bf 44} 
(1992) 43.}

\bibitem{adhm}{M.~Atiyah, V.~Drinfeld, N.~Hitchin and Y.~Manin, 
Phys. Lett. 
{\bf 65A} (1978) 185.}

\bibitem{aty}{M.~Atiyah, {\it Geometry of \YM fields}, Lezioni 
Fermiane, Accademia Nazionale dei Lincei e Scuola Normale 
Superiore, Pisa 
(1979).}

\bibitem{hitb}{N.~Hitchin, Math. Proc. Camb. Phil. Soc. 
{\bf 85} (1979) 465.}

\bibitem{gh}{G.~Gibbons, S.~Hawking, Phys. Lett. {\bf 78B} (1978) 430.}

\bibitem{eh}{T.~Eguchi and A.J.~Hanson, Ann.Phys.{\bf 120} (1979) 82.}

\bibitem{kro}{P.B.~Kronheimer, J. Diff. Geom. {\bf 29} (1989) 665.}

\bibitem{thooft}{G.~'t Hooft, Phys. Rev. {\bf D14} (1976) 3422; erratum, 
Phys. Rev. {\bf D18} (1978) 2199.}

\bibitem{jnr}{R.~Jackiw, C.~Nohl and C.~Rebbi, Phys. Rev. {\bf D15}
(1977) 1642.}

\bibitem{hklr}{N.J.~Hitchin, A.~Karlhede, U.~Lindstr\"om and 
M.~Ro\v cek, Comm. Math. Phys. {\bf 108} (1987) 535.}

\bibitem{mkay}{J.~McKay, Proc. Sympos. Pure Math., {\bf 37} (1980) 
183, A.M.S. Providence R.I.}

\bibitem{slan} {R.~Slanski, Phys. Rep. {\bf 79} (1981) 1.}

\bibitem{yui}{A.L.~Yuille, Class. Quant. Grav. {\bf 4} (1987) 1409.}

\bibitem{ver}{E.~Verlinde, Nucl. Phys. {\bf B455} (1995) 211.}

\bibitem{wittab}{E.~Witten, {On S-duality in Abelian Gauge Theory}, 
preprint IASSNS-HEP-95-36, hep-th/9505186.} 

\bibitem{gin}{For a review see P.~Ginsparg, {\it ``Applied Conformal Field 
Theory"}, in ``Fields, Strings and Critical Phenomena", Les Houches Summer 
School 1988, North Holland.}

\bibitem{vw}{C.~Vafa, E.~Witten, Nucl. Phys. {\bf B431} (1994) 3.}

\bibitem{bpst}{A.A.~Belavin, A.M.~Polyakov, A.S.~Schwarz and Yu.S.~Tyupkin,
Phys. Lett. {\bf 59B} (1975) 85.}

\bibitem{acto}{G.~'t Hooft, unpublished. 
For a review see A.~Actor, Rev. Mod. Phys.
{\bf 51} (1979) 461.}

\bibitem{cws}{N.~Christ, E.~Weinberg, N.~Stanton, Phys. Rev. {\bf D18}
(1978) 2013.}

\bibitem{pag}{D.~Page, Phys. Lett. {\bf 85B} (1979) 369.}

\bibitem{cg}{E.~Corrigan and P.~Goddard, Ann. Phys. {\bf 154} (1984) 253.}

\bibitem{gpr}{G.~Gibbons, C.~Pope and H.~R\"omer, Nucl. Phys. {\bf B157} 
(1979) 377.}

\bibitem{abfmr}{M.~Bianchi, F.~Fucito, M.~Martellini, G.C.~Rossi, Phys. Lett.
{\bf B359} (1995) 49.}

\bibitem{fin}{D.~Finkbeiner, {\it Introduction to Matrices and Linear 
Transformations}, Freeman and Company (1966).}

\bibitem{hita}{N.~Hitchin, {\it Metrics on Moduli Spaces}, in Proceedings of 
the Lefschez Centennial Conference (Contemporary Math. 58 Part I), 
A.M.S. Providence R.I. (1986).}

\bibitem{fu}{D.S.~Freed and K.K.~Uhlenbeck, {\it Instantons and 
Four-Manifolds},
MSRI Publ. 1, Springer-Verlag (1984).}

\bibitem{dk}{S.K.~Donaldson and P.B.~Kronheimer, {\it The 
Geometry of Four-Manifolds}, Oxford University Press (1990).}

\bibitem{yabe}{L.~Yaffe, Nucl. Phys. {\bf B151} (1979) 247;
C.~Bernard, Phys. Rev. {\bf D19} (1979) 3013.}

\end{thebibliography}
\end{document}